\documentclass{aa501}
\usepackage{amsmath,amssymb,epsfig}
\usepackage{natbib}

\newcommand{\rmd}{{\rm d}}
\newcommand{\rme}{{\rm e}}

\newcommand{\bx}{{\mathbf{x}}}
\newcommand{\by}{{\mathbf{y}}}
\newcommand{\bz}{{\mathbf{z}}}
\newcommand{\BE}{{\mathbb{E}}}
\newcommand{\BR}{{\mathbb{R}}}
\newcommand{\CA}{{\cal A}}

\newcommand{\CG}{{\cal G}}
\newcommand{\CP}{{\cal P}}
\newcommand{\CR}{{\cal R}}

\newcommand{\densM}{{m}}
\newcommand{\specM}{{\overline{M}}}
\newcommand{\reflex}{{\textsc{Reflex} }}
\newcommand{\hMpc}{{\ifmmode{h^{-1}{\rm Mpc}}\else{$h^{-1}$Mpc}\fi}}
\newcommand{\Beins}{\mbox{1\hspace*{-0.085cm}l}}

\begin{document}

%\thesaurus{(12.12.1; 12.03.3)}

\title{Non--Gaussian morphology on large scales:\\ 
Minkowski functionals of the  REFLEX cluster catalogue
}
\author{
M.~Kerscher\inst{1,2} \and 
K.~Mecke\inst{3,4} \and 
P.~Schuecker\inst{5} \and 
H.~B{\"o}hringer\inst{5} \and 
L.~Guzzo\inst{6} \and
C.~A.~Collins\inst{7} \and
S.~Schindler\inst{7} \and
S.~De~Grandi\inst{6} \and
R.~Cruddace\inst{8}
}
\institute{
Sektion Physik, Ludwig--Maximilians--Universit{\"a}t, 
Theresienstra{\ss}e 37, D--80333 M{\"u}nchen, Germany
\and
Department of Physics and Astronomy, 
The Johns Hopkins University, Baltimore, MD 21218, USA
\and
Max-Planck-Institut f\"ur Metallforschung, Heisenbergstr. 1, D-70569
Stuttgart, Germany   
\and
Institut f\"ur Theoretische und Angewandte Physik, Fakult\"at f\"ur
Physik, Universit\"at Stuttgart, Pfaffenwaldring 57, D-70569 Stuttgart, Germany
\and
Max-Planck-Institut f\"ur Extraterrestrische Physik, 
P.O.~Box: 1603, Giessenbachstrasse 1, D-85740 Garching, Germany
\and
Osservatorio Astronomico di Brera, Merate, Italy
\and
Liverpool John Moores University, Liverpool, UK
\and
Navel Research Laboratory, Washington DC, USA
}

\date{submitted May 9, 2001, revised July 17, 2001}

\titlerunning{Non--Gaussian morphology from the {}\reflex cluster catalogue}
\authorrunning{M.~Kerscher et al.}

%%%%%%%%%%%
%\begin{abstract}
\abstract{
In order to quantify higher--order correlations of the galaxy cluster
distribution we use a complete family of additive measures which give
scale--dependent morphological information. Minkowski functionals can
be expressed analytically in terms of integrals of $n$--point
correlation functions. They can be compared with measured Minkowski
functionals of volume limited samples extracted from the {}\reflex
survey.  We find significant non--Gaussian features in the
large--scale spatial distribution of galaxy clusters. A Gauss--Poisson
process can be excluded as a viable model for the distribution of
galaxy clusters at the significance level of 95\%.
 }
%\end{abstract}

\maketitle

%%%%%%%
\section{Introduction}

The spatial distribution of galaxy clusters poses important
constraints on cosmological models.  The abundance of clusters and
especially its evolution with redshift is very sensitive to parameters
of the cosmological models (see e.g.\
{}\citealt{kitayama:constraints}, {}\citealt{borgani:cosmological},
{}\citealt{bahcall:clusters},
{}\citealt{kerscher:abundance}).  To quantify the large--scale
structures traced by the galaxy clusters we have to go beyond the
number density.

Scenarios describing the formation of structures in the Universe start
with a mass density field showing only small deviations from the mean
density.  Inflationary scenarios suggest that these density
fluctuations can be modeled as a Gaussian random field completely
specified by its mean value and the power spectrum or two--point
correlation function (e.g.\ {}\citealt{kolb:early}).
In the initial stages of structure formation the linear approximation
is often used to evolve these fluctuations preserving their Gaussian
nature and increasing their amplitude only (see e.g.\
{}\citealt{peebles:lss}).  With growing over--density the nonlinear
couplings become more and more important leading to a non--Gaussian
density field.  Also the process of galaxy formation may introduce
non--Gaussian features if the ``biasing'' is non--linear (see e.g.\
{}\citealt{scoccimarro:bispectrum}).  Typically one argues that on
large scales, the evolution is still in the linear regime, and one
expects that the smoothed density field is proportional to the initial
Gaussian field.  However, structures like walls and filaments were
observed in the galaxy distribution on large scales
{}\citep{huchra:cfa2s1,shectman:lcrs}.  These non--Gaussian features
appear at a low density contrast and are therefore hard to detect.
The sensitivity of the Minkowski functionals, even if only a small
number of points is available, allows us to quantify the non--Gaussian
morphology of these structure on large scales.  Walls and filaments
were predicted by analytical and numerical work based on the
Zel'dovich approximation
{}\citep{zeldovich:fragmentation,arnold:largescale,doroshkevich:superlargeII}
and related approximations {}\citep{kofman:coherent,bond:filament}.
$N$--body simulations could verify that these structures are generic
features of the gravitational collapse for Cold Dark Matter~(CDM) like
initial conditions {}\citep{melott:generation,jenkins:evolution}.

Since observations supply us with the positions of galaxies and galaxy
clusters in space, our methods will use this point distribution
directly. No smoothing is involved.  Therefore we have to give a clear
definition what a ``Gaussian'' point distribution, the Gauss--Poisson
process, is.  Some of the statistical properties of random fields
directly translate to similar statistical properties of point
distributions, but also important differences show up.  The
equivalence of the Gauss--Poisson process with a simple Poisson
cluster process, allows us to simulate a `Gaussian'' point
distribution {}\citep{kerscher:constructing}. With these simulations
we will perform a Monte--Carlo test to determine the significance of
the non--Gaussian features in the cluster distribution.

Statistical measures provide important tools for the comparison of the
large--scale structure in the Universe with theoretical models.  The
discriminative power of this comparison depends chiefly on the
statistical measure.  The most frequently employed measure was and
still is the two--point correlation function, or the power spectrum.
Both are nowadays an imperative in the analysis of any galaxy or
cluster catalogue: for the {}\reflex cluster catalogue see
{}\citet{collins:spatial} and {}\citet{schuecker:reflexIII}.  They
give important information on the fluctuation spectrum of matter.
However, they appear to be blind to morphological features.  Indeed,
completely different spatial patterns and point distributions could
display the same two--point correlation function, i.e., no direct
conclusions about the morphology of the structure can be drawn from an
analysis with these two--point measures
{}\citep{baddeley:cautionary,szalay:walls,jun:largescale,kerscher:constructing}.
Higher--order correlation functions immediately come to mind if one
wants to go beyond the two--point correlation function.  And indeed
three--point correlations were detected in the distribution of galaxy
clusters {}\citep{toth:threepoint}.  However, there is a conceptual
problem since $n$--point functions ($n\ge3)$ depend on $3(n-1)-3$
parameters even for isotropic and homogeneous point distributions.
Already for the three--point correlation function we are not aware of
a study where its dependence on all three parameters was
estimated. Clearly, integral information is mandatory and necessary.
This may be accomplished e.g.\ for the three--point function by
averaging over the shape of triangles, or by considering the
(factorial) moments of counts in cells (see e.g.\
{}\citealt{peebles:lss} and {}\citealt{szapudi:higher}).  Another
effort to go beyond the two--point correlation function comprises the
percolation analysis {}\citep{shandarin:percolation}. Also the genus,
closely related to the Euler characteristic, is often employed to
quantify deviations from a Gaussian density fields (see e.g.\
{}\citealt{hamilton:topology}, {}\citealt{melott:review} and
references therein).

For the construction of statistical methods, sensitive to the 
large--scale structures, additivity is a heuristic principle which can 
guide us how to define useful measures which do not depend on all 
these parameters.  Additivity yields robust, local decomposable 
measures.  The mathematical discipline of integral geometry (see e.g.\ 
{}\citealt{hadwiger:vorlesung}) supplies us with a suitable family of 
such descriptors, known as Minkowski functionals.  These measures 
embody information from every order of the correlation functions, are 
numerically robust even for small samples, and yield global as well as 
local morphological information.  The Minkowski functionals are 
additive measures which allows us to calculate them efficiently by 
summing up their local contributions although they depend on all 
orders of correlation functions.
The application of Minkowski functionals in statistical physics and 
cosmology are reviewed by {}\citet{mecke:additivity} and 
{}\citet{kerscher:statistical}, respectively.

Samples of galaxy clusters are based mainly on optical observations,
where the clusters are selected as galaxy over--densities in the
two--dimensional maps on the celestial sphere (c.f.\
{}\citealt{abell:distribution}, {}\citealt{abell:catalog},
{}\citealt{dalton:apmcluster}, and {}\citealt{gal:northern}).
Projection effects seem to have a non--negligible effect on the
statistical analysis of these optically selected cluster samples
{}\citep{katgert:enacs-i,haarlem:projection}.  Only in recent years
X--ray selected cluster samples have been completed.  Since the X--ray
luminosity is proportional to the baryonic density squared,
over--densities are more emphasized.  Consequently, the contamination of
the catalogue by chance alignments due to projections is reduced
{}\citep{boehringer:reflexI}. Assuming a virial relation, the X--ray
luminosity of the galaxy cluster can be related to its mass.

%%%%%%%
\section{Morphology of large scale structure}

Minkowski functionals have been introduced to cosmology as a tool to
quantify the morphology of large--scale structures by
{}\citet{mecke:robust} where also a first analysis of the distribution
of galaxy clusters based on the {}\citet{abell:catalog} sample with a
redshift compilation by {}\cite{postman:distribution} was presented.

With Minkowski functionals we quantify the morphology of a
sufficiently well behaved compact body $K\subset\BR^3$ by assigning it
a number $M_\nu(K)\in\BR$.  The Minkowski functionals are motion
invariant
\begin{equation} 
M_\nu(gK)=M_\nu(K),  
\end{equation} 
where $g=(\bx,\mathbf{\Theta})$ are the movements in three dimensions,
i.e. translations $\bx$ and rotations $\mathbf{\Theta}$.  As already
emphasized the additivity property
\begin{equation} 
\label{eq:additivity}
M_\nu(K\cup K')=M_\nu(K)+M_\nu(K)- M_\nu(K\cap K') 
\end{equation}
serves as the construction principle of these measures.
All the Minkowski functionals have a straightforward interpretation in
terms  of  geometrical and  topological  quantities  as summarized  in
Table~\ref{table:mingeom}.
\begin{table}
\caption{\label{table:mingeom} Minkowski functionals $M_\mu$ and the
normalized Minkowski functionals $\Phi_\mu$ (Eq.~\eqref{eq:def-Phi})
in three--dimensional space expressed in terms of more familiar
geometric quantities.}
\begin{center}
\begin{tabular}{lc|c|c|c}
geometric quantity      &        & $\mu$ & $M_\mu$        & $\Phi_\mu$ \\[1ex]
\hline
volume                  & $V$    & 0 	 & $V$ & $V/(\tfrac{4\pi}{3}r^3N)$ \\
surface area            & $A$    & 1 	 & $A/8$          & $A/(4\pi r^2N)$ \\
integral mean curvature & $H$    & 2 	 & $H/(2\pi^2)$   & $H/(4\pi rN)$ \\
Euler characteristic    & $\chi$ & 3 	 & $3\chi/(4\pi)$ & $\chi/N$ \\
\end{tabular}
\end{center}
\end{table}
Minkowski functionals are distinguished from other geometric measures
by the theorem of {}\citet{hadwiger:vorlesung}, which states that
there are only four independent scalar functionals in
three--dimensional space, which are motion invariant, additive, and
continuous for convex bodies.  Hence, every additive geometrical
measure $M(K)$ which does not depend on the position and orientation
of the body $K$ in three--dimensional space, can be written as a
linear combination of the four Minkowski functionals:
\begin{equation}
M(K) = \sum_{\nu=0}^3 c_\nu M_\nu(K) .
\end{equation}
{}\citet{beisbart:morphometry} discuss the vector valued extensions of
the Minkowski functionals.

\begin{figure}
\begin{center}
\epsfig{file=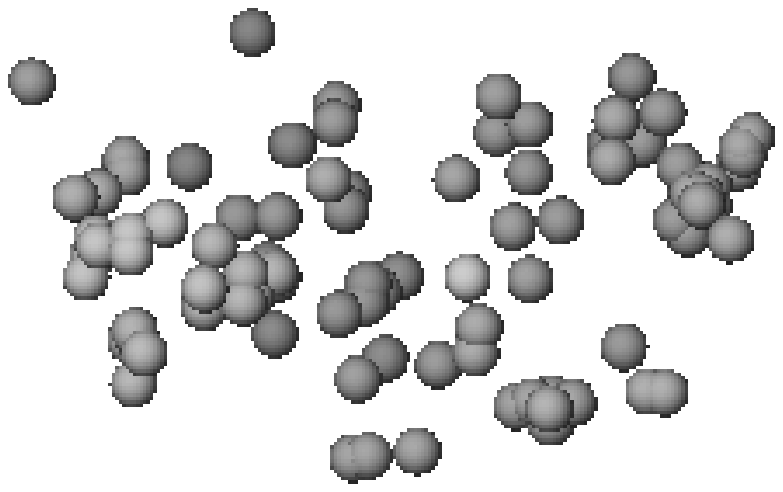,width=6cm}
\epsfig{file=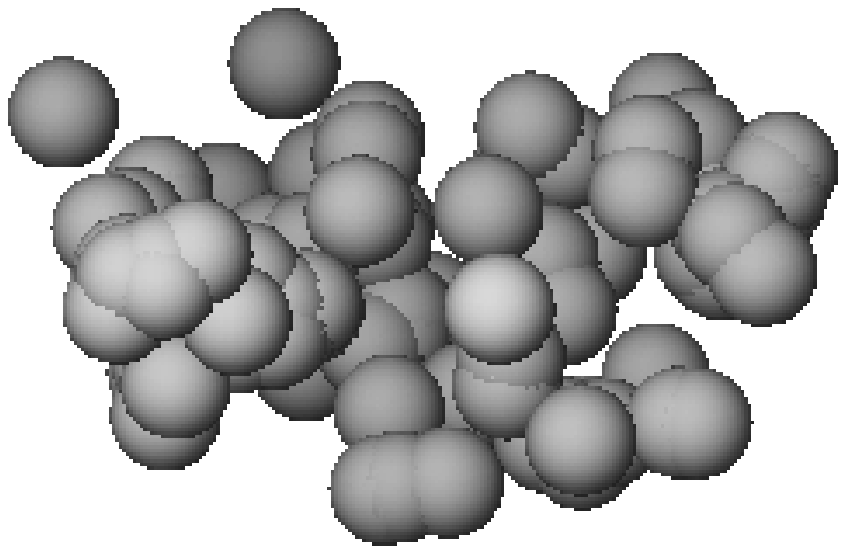,width=6cm}
\epsfig{file=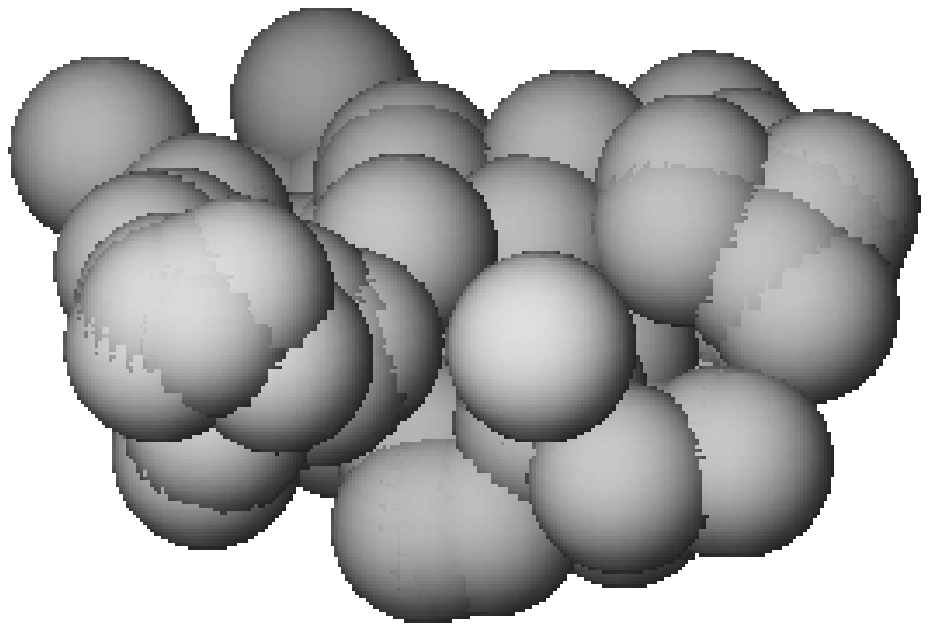,width=6cm}
\end{center}
\caption{\label{fig:bobbel} Dilated points: spheres of varying radius
attached to the galaxy cluster from the {}\reflex sample L12 (see
Table~\ref{table:samples}).}
\end{figure}

The cluster distribution provides us with a point set
$X=\{\bx_i\}_{i=1}^N$ in three--dimensional space.  One may think of
$X$ as a skeleton of the large--scale structures in the Universe.
Since there are only four numbers of the MFs in three dimensions,
compared to a correlation function, it is necessary to define
morphological functions $M_\mu(r)$.  The proper technique to do this
for general random spatial structures are (erosion) dilation
operations (see Fig.~\ref{fig:bobbel}, and {}\citealt{serra:image}).
In case of point patterns this techniques reduces to fixing balls
$B_r$ of radius $r$ at each point.  With Minkowski functionals we
quantify the geometry and topology of union set of these balls
$\CA_r=\bigcup_{i=1}^N B_r(\bx_i)$.  The radius $r$ is employed as a
diagnostic parameter.  In such a way, we obtain scale--dependent
integral information on higher--order correlations of the distribution
of galaxy clusters and not only two--point correlation functions of
the large--scale distribution.
Erosion/dilation techniques combined with additive Minkowski
functionals have been successfully applied in many areas, including
condensed matter physics {}\citep{mecke:additivity}, geology
{}\citep{arns:characterization,arns:predicting}, and
digital image analysis {}\citep{serra:image,serra:imageII}.

\subsection{The Boolean grain model}
\label{sect:booleangrain}

The simplest model of a random point distribution is the Poisson
process.  By attaching grains to each of the points, in our case balls
of radius $r$, we arrive at the Boolean grain model.
{}\citet{mecke:euler} presented a method to calculate mean volume
densities of the Minkowski functionals for this model.  We repeat
their arguments, since its extension allows us to calculate the
Minkowski functionals of correlated grains in
Sect.~\ref{sect:MF-correlated}.

Iterating the additivity relation {}\eqref{eq:additivity} one obtains
for the union $\CA_r=\bigcup_{i=1}^N B_r(\bx_i)$ of $N$ spheres
$B_i=B_r(\bx_i)$ of radius $r$ and center $\bx_i$ the
inclusion--exclusion formula
\begin{multline}
\label{eq:inclusion-exclusion}
M_{\nu}(\CA_r) = \sum_i M_{\nu} (B_i) 
- \sum_{i<j} M_{\nu} (B_i \cap B_j) +\\
+ \ldots (-1)^{N+1} M_{\nu} (B_1 \cap \ldots \cap B_N )\;. 
\end{multline} 
Generally, a point process in a domain $\Omega $ with volume
$|\Omega|$ is specified by a sequence of product densities
$\varrho_n(\bx_1,\ldots,\bx_n)$ with the mean number density
$\varrho\equiv\varrho_1(\bx_1)$. $\varrho_n(\bx_1,\ldots,\bx_n)\rmd
V_1\ldots\rmd V_n$ is the probability of finding $n$ points
in the volume elements $\rmd V_1$ to $\rmd V_n$.  The volume density
$\densM_{\nu}(\CA_r)$ of the $\nu$--th Minkowski functional per unit
volume for the augmented coverage $\CA_r$ are then obtained from the
inclusion--exclusion formula~\eqref{eq:inclusion-exclusion} in the form
\begin{multline} 
\label{eq:corradd}
\densM_{\nu}\big(\CA_r;\{\varrho_n\}\big) = \\
\sum_{n=1}^\infty \frac{(-1)^{n+1}}{n! |\Omega|}
\int\limits_\Omega \rmd\Gamma_n\ M_\nu
\Bigg( \bigcap_{i=1}^n B_r(\bx_i)\Bigg) \ \varrho_n(\Gamma_n) 
\end{multline}
where we introduced, for convenience, the variable
$\Gamma_n=(\bx_1,\ldots,\bx_n)$ with the integration measure
$\int_{\Omega}d\Gamma_n=\prod_{i=1}^n \int_{\Omega}d\bx_i$.
Obviously, the Minkowski functionals embody information from every
order $n$ of the $n$--point densities $\varrho_n(\bx_1,\ldots,\bx_n)$.
If the product densities $\varrho_n(\bx_1,\ldots,\bx_n)=\varrho^n$
were independent of position (Poisson distribution of density
$\varrho=N/|\Omega|$), the integrals in Eq.~\eqref{eq:corradd} can be
performed using the fundamental kinematic formula
({}\citealt{blaschke:I}, {}\citealt{santalo:integralgeometry})
\begin{equation} 
\label{eq:kinematic}
\int\limits_\CG\rmd g\ M_{\nu}(K \cap gK') =
\sum_{\mu=0}^{\nu}\binom{\nu}{\mu} M_{\mu}(K) M_{\nu-\mu}(K') .    
\end{equation}  
Equation~\eqref{eq:kinematic} describes the factorization of the
Minkowski functionals of the intersection $K\cap K'$ of two bodies $K$
and $K'$ if one integrates over the motions $g=(\bx,\mathbf{\Theta})$,
i.e.  translations $\bx$ and rotations $\mathbf{\Theta}$ of $K'$.  For
Poisson distributed spheres $B_r$ of radius $r$ one obtains the mean
values of the Minkowski functionals per unit volume
(\citealt{mecke:euler}, {}\citealt{stoyan:stochgeom})
\begin{equation} 
\label{eq:mean1} 
\densM_\nu(\CA_r;\varrho) =  \frac{\partial^\nu}{\partial t^\nu}
\Bigg\{1-\exp\Bigg[-\varrho \sum\limits_{\mu=0}^d \frac{t^\mu}{\mu!} M_\mu(B_r)
\Bigg]\Bigg\} \Bigg|_{t=0}  
\end{equation}  
and  particularly   in  three  dimensions   the  normalized  Minkowski
functionals
\begin{equation}
\label{eq:def-Phi}
\Phi_\nu(r) = \frac{\densM_\nu(\CA_r)}{M_\nu(B_r)\; \varrho} 
\end{equation}
read with $\eta=\frac{4\pi}{3}r^3\varrho$:
\begin{equation} 
\label{eq:mean3d}
\begin{array}{ll}
\Phi_0(r;\varrho) = &   (1-\rme^{-\eta})/\eta \;\;, \\
\Phi_1(r;\varrho) = &   \rme^{-\eta} \;\;, \\
\Phi_2(r;\varrho) = &  (1-\frac{3\pi^2}{32} \eta) \rme^{-\eta}\;\;,\\
\Phi_3(r;\varrho) = &  (1-3\eta+\frac{3\pi^2}{32} \eta^2) \rme^{-\eta}\;\;.  \\
\end{array} 
\end{equation}

%%%
\subsection{The {}\reflex cluster sample}

The construction of the {}\reflex cluster sample is described in
detail by {}\citet{boehringer:reflexI} and the statistics of the
cluster distribution is described by {}\citet{collins:spatial} for the
two-point correlation function and in {}\citet{schuecker:reflexIII}
for the density fluctuation power spectrum.  The survey area covers
the southern sky up to the declination $\delta\le2.5^o$ avoiding the
band of the Milky Way, $|b_{II}|\le20^o$ and the regions of the
Magellanic clouds.  The total survey area is 13924~$\deg^2$ or
4.24~ster.  Tests including the number counts
($\log{N}\log{S}$--function), the comoving densities, $\langle
V/V_{max}\rangle$ tests, and comparisons to simulations, described in
the above mentioned papers, show that the selection function is well
documented.

The X--ray detection of the clusters is based on the second processing
of the RASS (ROSAT All Sky Survey, {}\citealt{voges:bcs}) exploiting a
primary (MPE internal) source detection list comprising 54076 sources
in the {}\reflex area down to a detection likelihood of $L\ge7$ (see
{}\citealt{voges:bcs}). For all these sources the X--ray parameters
are reanalysed by the growth curve analysis method as described by
{}\citet{boehringer:northern} which provides a flux measurement with
significantly less discrimination against extended X--ray sources than
provided by the standard analysis of the RASS. The results of this
reanalysis are used to produce a flux--limited sample of RASS sources
with a nominal flux $F_x\ge3\cdot10^{-12}$ erg s$^{-1}$ cm$^{-2}$ in
the energy band (0.1--2.4~keV).

The cluster candidates are finally identified or removed from the
sample as non--cluster sources by a detailed documentation of the
X--ray and optical source properties, literature information, and
spectroscopic information including redshift measurements obtained by
follow--up observations within the frame of an ESO key
program. Further tests of the sample completeness based on a search
for clusters among the significantly extended X--ray sources and a
search for X--ray emission from clusters cataloged by
{}\citet{abell:catalog} independent of the RASS source detection
supports the completeness estimate of $>90\%$ for a flux--limit of
$3\cdot10^{-12}\text{erg}\text{s}^{-1}\text{cm}^{-2}$. The high
completeness concerning the optical identification makes the data set
an effectively X--ray selected sample of galaxy clusters.  The final
cluster sample includes 452 clusters and there are three objects left
in the list with uncertain identifications and redshifts. These three
objects are excluded here in the further analysis.

\begin{figure}
\begin{center}
\epsfig{file=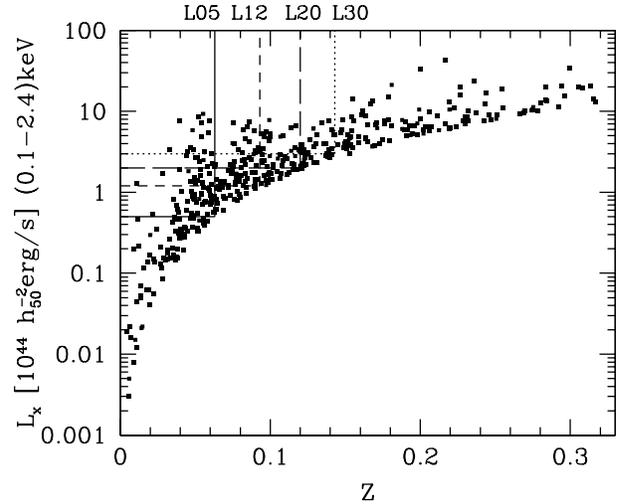,width=8cm}
\end{center}
\caption{\label{fig:volume-limit} Luminosity -- redshift distribution
of {}\reflex clusters of galaxies (points) and the applied ranges for
the extraction of volume--limited subsamples. The redshift and
luminosity intervals of the respective volume--limited subsamples L05,
L12, L20, and L30 are marked by continuous, short--dashed, long--dashed,
and dotted lines. The subsamples are described in
Table~\ref{table:samples}. Note that for conventional reasons the X--ray
luminosities are given in units of $H_0=50\,{\rm km}\,{\rm
s}^{-1}\,{\rm Mpc}^{-1}$.}
\end{figure}

For the determination of MFs complete volume--limited subsamples are
needed. The {}\reflex cluster sample is per construction X--ray
flux--limited so that the fraction of luminous clusters increases with
redshift (see Fig.~\ref{fig:volume-limit}). Volume--limited
distributions are selected by introducing upper redshift and lower
luminosity limits (vertical and horizontal lines in
Fig.~\ref{fig:volume-limit}). In order to reduce possible (error
migration) effects which might occur at the flux--limit (e.g.,
{}\citealt{eddington:correction}) the upper redshift limits are set
slightly below the formal redshift limit, especially for large
luminosities where the effects could be largest.

The completeness of the different volume--limited subsamples is
illustrated in Fig.\,3 in {}\citet{schuecker:reflexIII} which includes
the subsamples denoted by L05 to L30 in
Table~\ref{table:samples}. Similar volume--limited samples as listed in
Table~\ref{table:samples} have been used by {}\citet{collins:spatial}
in their analysis using the two--point correlation function.  Comoving
distances have been calculated according to the Mattig formula with
$\Omega=1$, $h=0.5$, and $\Lambda=0$.  The flat redshift--independent
distribution of comoving cluster number density suggests the absence
of large incompleteness effects of the subsamples in the redshift
range $0\le{z}\le0.15$. We thus expect no significant artificial
fluctuations introduced by incompleteness effects on scales up to
comoving radial distances of $R<400\,h^{-1}\,{\rm Mpc}$.

\begin{table}
\begin{center}
\caption{ \label{table:samples} The volume--limited cluster samples 
used in our investigations consisting out of $N$ clusters closer than 
$R$ with X--ray luminosity higher than $L_{\rm min}$.  The constraint 
$\varrho C_2$ will be explained in Sect.~\ref{sect:gauss-cluster}.} 
\vspace{0.2cm}
\begin{tabular}{c|c|c|c|c|c}
Sample & $R$     & $L_{\rm min}$     & N  & $\varrho$ & $\varrho C_2$ \\ 
       & [Mpc/h] & [erg/s] &    & [$h^3\text{Mpc}^{-3}$]& \\[1ex]
\hline
L05 & 180 & $0.5\times10^{44}$ & 74 & $8.9\times10^{-6}$ & 3.2 \\
L12 & 260 & $1.2\times10^{44}$ & 95 & $3.8\times10^{-6}$ & 2.2 \\
L20 & 330 & $2.0\times10^{44}$ & 86 & $1.7\times10^{-6}$ & 0.6 \\
L30 & 385 & $3.0\times10^{44}$ & 62 & $0.8\times10^{-6}$ & 0.3 \\
\end{tabular}
\end{center}
\end{table}

\subsection{Minkowski functionals of the {}\reflex clusters}
\label{sect:minkowski-reflex}

\begin{figure}
\begin{center}
\epsfig{file=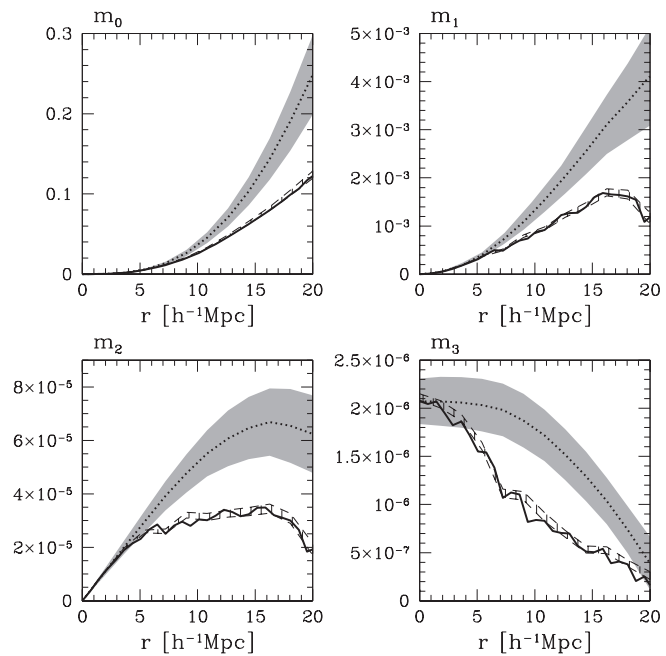,width=9cm}
\end{center}
\caption{\label{fig:min05} Minkowski functionals of the volume--limited
sample L05 (solid line) compared to the Minkowski functionals of a
Poisson process with the same number density (dotted line, gray shaded
one--$\sigma$ area).  The numerically determined mean is in perfect
accordance with Eq.~\eqref{eq:mean3d}.  The dashed one--$\sigma$ area
is obtained by filling the excluded area around the small and large
Magelanic clouds with randomly distributed points of the same number
density as the rest of the sample.}
\end{figure}

\begin{figure}
\begin{center}
\epsfig{file=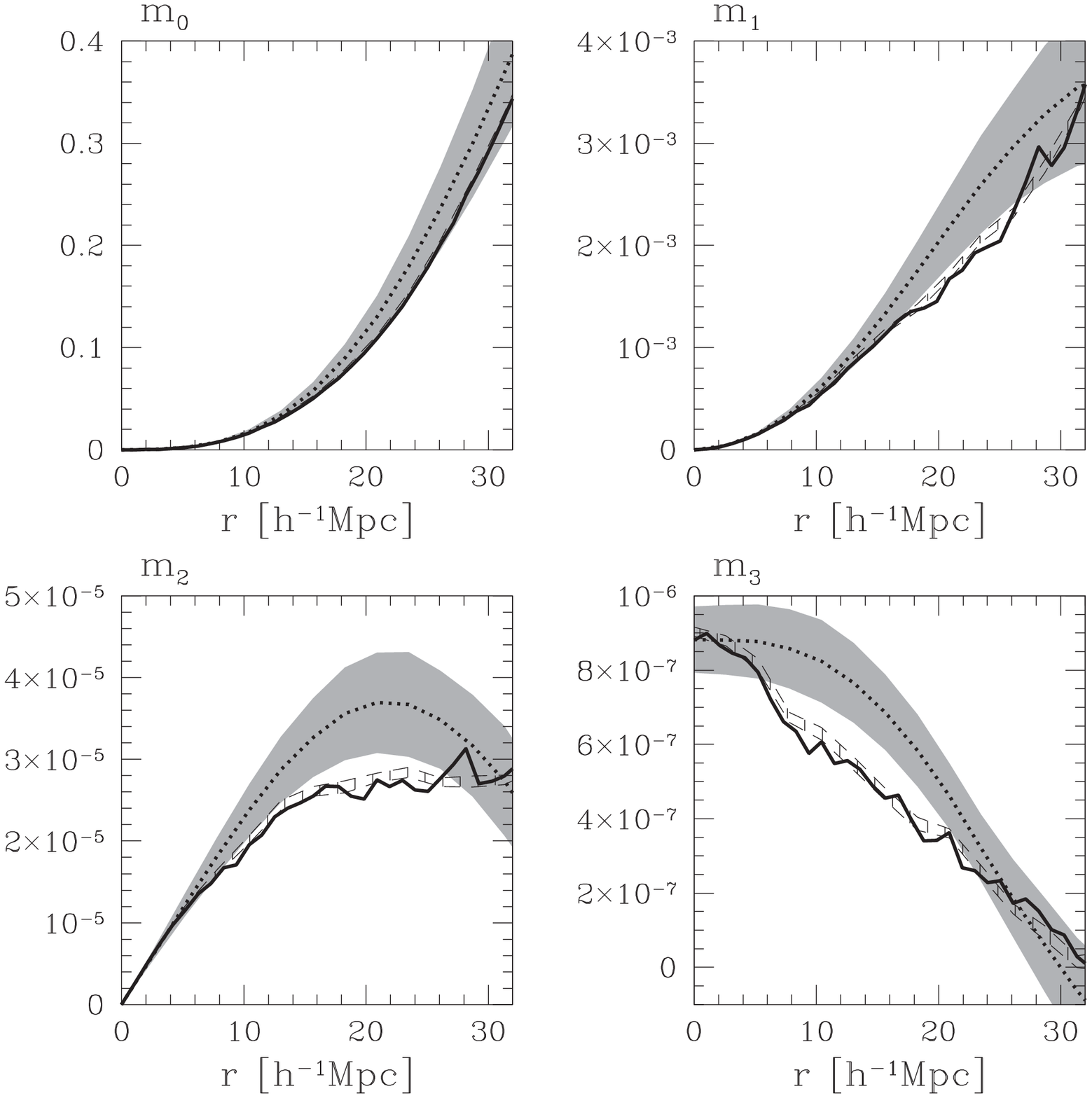,width=9cm}
\end{center}
\caption{\label{fig:min12} Minkowski functionals of the volume--limited
sample L12. Same conventions as in Fig.~\ref{fig:min05}.}
\end{figure}

\begin{figure}
\begin{center}
\epsfig{file=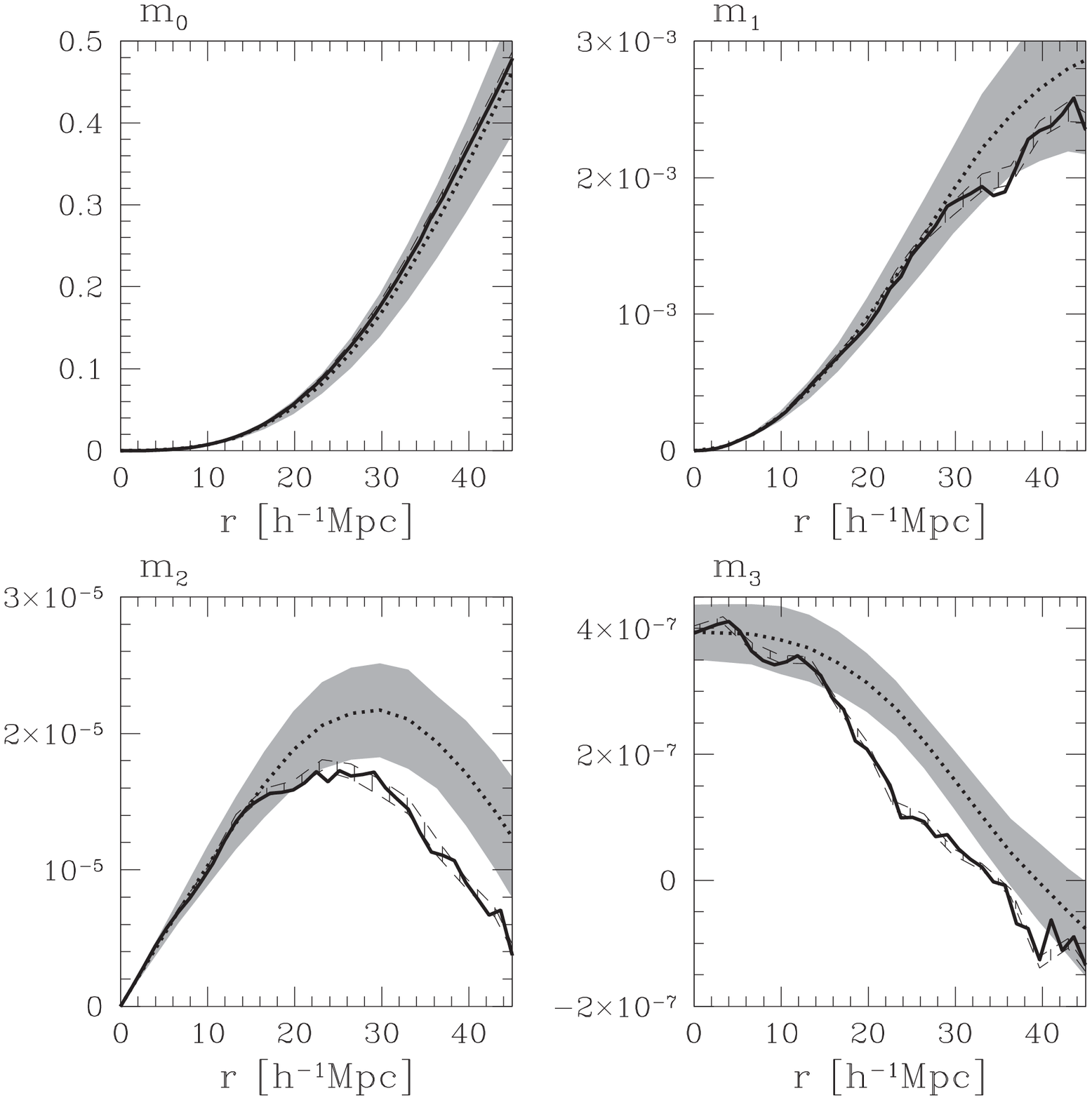,width=9cm}
\end{center}
\caption{\label{fig:min20} Minkowski functionals of the volume--limited
sample L20. Same conventions as in Fig.~\ref{fig:min05}.}
\end{figure}

\begin{figure}
\begin{center}
\epsfig{file=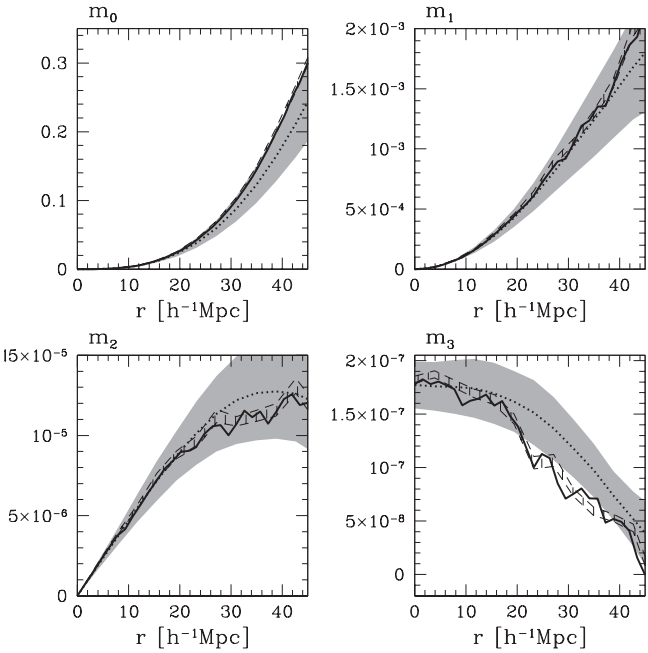,width=9cm}
\end{center}
\caption{\label{fig:min30} Minkowski functionals of the volume--limited
sample L30. Same conventions as in Fig.~\ref{fig:min05}.}
\end{figure}

To study the morphology of the large--scale distribution of galaxy
clusters we consider a series of volume--limited samples from the
{}\reflex cluster catalogue {}\citep{boehringer:reflexI}.
The volume densities $\densM_\mu$ of the Minkowski functionals were
calculated using the minus--sampling boundary correction, based on
partial Minkowski functionals as suggested by {}\citet{mecke:robust}
(for details see {}\citealt{schmalzing:cfa2}).
The survey is bounded by $\delta<2.5^\circ$ and $|b|>20^\circ$, but
also several regions in the small and large Magelanic clouds were
excluded from the sample (for details see
{}\citealt{boehringer:reflexI}).  To estimate their influence on the
Minkowski functionals of the samples we filled these regions with
random points with the same number density.  The comparison in
Figs.~\ref{fig:min05}--\ref{fig:min30} shows nearly identical results
for filled or unfilled regions in the Magelanic clouds.


The overall features seen in the Minkowski functionals of the \reflex 
clusters are similar to the one observed in the Abell/ACO cluster 
sample of {}\citet{plionis:evidence} as analyzed by 
{}\citet{kerscher:abell}.  We are limited by the smaller sample size 
and the boundary correction used.  Only few galaxy clusters contribute 
to the Minkowski functionals for large radii.  Therefore we are not 
able to trace the large--scale structure to the limit where the sample 
volume is filled by the union set of balls.
Additionally to the Minkowski functionals of the clusters the results
for a Poisson process with the same number density inside the sample
geometry is shown in the Figs.~\ref{fig:min05}--\ref{fig:min30}.
Increasing the depth of the volume--limited samples from L05 to L30 the
Minkowski functionals show a clear trend from strong clustering
towards only small differences from the Poisson distribution.
Increasing the depth of the volume--limited samples we allow for
clusters with higher X--ray luminosity.  Considering the amplitude of
the two--point correlation function, galaxy clusters with higher
X--ray luminosity should show stronger correlations
{}\citep{kaiser:onspatial,bardeen:gauss}.  However, in the deeper
volume--limited samples, with the more luminous clusters, also the
number density decreases. The sparseness of the point distribution
competes with the increased amplitude of the two--point correlation
function.  Indeed, for quite general conditions, a point distribution
converges towards a Poisson process under thinning, i.e.\ under
randomly deleting points (e.g.\ {}\citealt{daley:introduction},
Sect.~9.3).
For the Minkowski functionals the behavior in sparse samples may be
explained by considering the expansion of the normalized Minkowski
functionals $\Phi_\mu$ in terms of $\eta=\varrho(4\pi/3)r^3$ around
zero (see also {}\citealt{kerscher:pscz}).  Based on the expansion
{}\eqref{eq:corradd} of MFs in terms of $n$--point densities one gets
to the lowest order in $\eta$
\begin{multline}
\label{eq:eta-expansion}
\Phi_\mu(\CA_r) = \\
1-\eta\ \frac{3}{2r^3}\int_0^{2r}\rmd s\ s^2\ I_\mu(r;s) 
\Big(1+\xi_2(s)\Big) + O(\eta^2) \;.
\end{multline}
The functions
\begin{equation}
\label{eq:define-Imu-3d}
\begin{array}{rl} 
I_0(r;s) =  &  1 - \frac{3}{2} \frac{s}{2r}  + 
\frac{1}{2}\big(\frac{s}{2r}\big)^3 ,\\    
I_1(r;s) = & 1 - \frac{s}{2r} ,\\ 
I_2(r;s) = & 1 - \frac{s}{2r} + 
 \big(\frac{\pi}{4} - \frac{1}{2} \arccos\frac{s}{2r} \big)
\Big(1-\big(\frac{s}{2r}\big)^2\Big)^\frac{1}{2} ,\\
I_3(r;s) = & \Theta(2r-s)\\ 
\end{array} 
\end{equation}
resemble the Minkowski functionals
$I_\nu(r;|\bx|)=M_\mu(B_r(\mathbf{0})\cap B_r(\bx))/M_\mu(B_r)$ of the
intersection of two spheres of radius $r$ and distance $|\bx|$.
$\Theta(q)$ is the step function equal zero for $q<0$ and one for
$q>0$.
In Eq.~\eqref{eq:eta-expansion} the integral and its pre--factor give a 
dimensionless geometric number depending on the correlation function 
$\xi_2(s)$ and the measure $M_\mu$.  Terms proportional to $\eta^n$ 
include intersections of $n$ spheres weighted by the $n$--point 
densities.  Only the two--point correlation function is important for 
small $\eta$, but the Minkowski functionals become increasingly more 
sensitive to higher--order correlations with larger $\eta$.  For very 
small $\eta$ we essentially arrive at a Poisson process as numerically 
verified by {}\citet{kerscher:fluctuations}.  A small $\eta$ may be 
obtained either with a small radius $r$ of the spheres, or, as in our 
case, by a low number density $\varrho$.

Now let us describe the features in the MFs in more detail.  The strong 
clumping in the distribution of galaxy clusters is causing the lowered 
values of the volume densities $m_\mu$ of the Minkowski functionals 
compared to the Poisson values.  In a clustered point distribution, 
the spheres in the union set $\CA_r=\bigcup_{i=1}^N B_r(\bx_i)$ 
overlap significantly already for small radii.  This is leading to a 
reduced density of the volume $m_0$, surface area $m_1$, and integral 
mean curvature $m_2$.  The density of the Euler characteristic $m_3$ 
decreases since for small radii mainly the number of connected objects 
is counted -- no tunnels and cavities have formed yet.  A tunnel 
through the body $\CA_r$ gives a negative contribution of minus one to 
the Euler characteristic.  In the sample L20 we observe the zero 
crossing of the Euler characteristic indicating that an interconnected 
network of tunnels, a sponge--like, bi--continuous ``cosmic web'' has 
formed for radii around 35\hMpc.  

In the deeper samples L20 and L30 the volume density $m_0$ shows a 
tendency towards increased values compared to a Poisson process.  With 
our estimator for the MFs, we successively shrink the sample 
proportional to $r$, where $r$ is the radius of the spheres $B_r$.  
Therefore, we mainly probe the central region of the sample for large 
$r$.  The increased $m_0$ is caused by gradients in the number density 
$\varrho$ of the {}\reflex cluster sample, specifically the local 
under--density of clusters out to approximately 100\hMpc\ (see 
{}\citealt{schuecker:reflexIII}, and for galaxies 
{}\citealt{zucca:esp-ii}).

There is no easy relation between the scale $s$ of fluctuations in the
number density as probed by $\xi_2(s)$ and the radius $r$ of the
spheres used in the MF analysis. As can be seen from
Eqs.~\eqref{eq:corradd} and {}\eqref{eq:meancorr3} weighted integrals
over all scales contribute to the MFs at a given radius $r$. 
However, with the radius $r$ of the spheres $B_r$ we probe the
geometry and topology of the cluster distribution in a
scale--dependent way. The radius $r$ can be regarded as a geometrical
scale, e.g.\ the radius $r_p$ of the first zero of the Euler
characteristic $m_3(\CA_{r_p})=0$ is an estimate of the percolation
threshold for our system of mono--disperse spheres
{}\citep{mecke:euler}. At this scale $r_p$ the large--scale structure
elements (walls, filaments, clusters) form a percolating network.

%%%%%%
\section{Higher--order correlations in point distribution}

%%%
\subsection{Minkowski functionals for correlated grains}
\label{sect:MF-correlated}

As for the Poisson process ({}\citealt{mecke:euler}, and
Sect.~\ref{sect:booleangrain}) one may calculate the Minkowski
functionals of correlated grains, in our case spheres centered on a
clustered point set ({}\citealt{mecke:diss},
{}\citealt{schmalzing:quantifying}, {}\citealt{mecke:additivity}).
Our main analytic result Eqs.~\eqref{eq:meancorr3} and
{}\eqref{eq:meancorr4} expresses the Minkowski functions $M_\nu(r)$ in
terms of centered correlation functions which allows a direct
comparison of measured functions with a Gaussian model where higher
correlations are set to zero.

The expression~\eqref{eq:corradd} may be used to calculate the
Minkowski functionals for correlated grains given the $n$--point
densities $\varrho_n(\bx_1,\ldots,\bx_n)$ of the point distribution.
An alternative and sometimes more convenient expression
for the densities $\densM_{\nu}(\CA_r)$ than Eq.~\eqref{eq:corradd}
can be obtained in terms of the cumulants, the {\em connected} or {\em
centered} correlation functions $\xi_n(\Gamma_n)$ with
$\xi_1(\bx_1)=1$. For the two--point correlation function we have
\begin{equation}
\xi_2(r)+1 = \xi_2(\bx_1,\bx_2)+1= \frac{\varrho_2(\bx_1,\bx_2)}{\varrho^2},
\end{equation} 
with $r=|\bx_2-\bx_1|$, and in general   
\begin{equation}
\varrho_n(\Gamma_n)= 
\varrho^n \sum_{\{\CP\}} \prod_{i=1}^{|\CP|}\xi_{p_i}(\Gamma_{p_i}) 
\end{equation}
Thus, the product densities $\varrho_n$ of order $n$ is given by a sum
over all possible partitions $\CP$ of the coordinates
$\Gamma_n=(\bx_1,\ldots,\bx_n)$ into $|\CP|$ parts of $p_i$ elements.
Each vector $\bx_j\in\Gamma_n$ occurs exactly once as an argument
$\bx_j\in\Gamma_{p_i}$ of a cumulant $\xi_{p_i}(\Gamma_{p_i})$ on the
right side, i.e., $\sum_{i=1}^{|\CP|}p_i=n$.

Using the additivity  relation~\eqref{eq:additivity} and the kinematic
formula~\eqref{eq:kinematic}  of  the  Minkowski functionals  one  can
follow the derivation in {}\citet{mecke:euler} so that one immediately
obtains the expression for the intensities {}\citep{mecke:diss}
\begin{multline}
\densM_{\nu}\big(\CA_r;\{\xi_n\}\big)= \\
\frac{\partial^\nu}{\partial t^\nu} 
\Bigg\{1-\exp\bigg[-\varrho \sum\limits_{\mu=0}^d \frac{t^\mu}{\mu!} 
\specM_\mu(r;\varrho,\{\xi_n\}) M_\mu(B_r) \bigg]\Bigg\} \Bigg|_{t=0}  
\end{multline} 
due  to the  factorization of  the integral  in Eq.~\eqref{eq:corradd}
(compare  with Eq.~\eqref{eq:mean1}).   The normalized  {\em specific}
Minkowski functionals
\begin{multline} 
\label{eq:meancorr3}
\specM_\nu(r;\varrho,\{\xi_n\}) = 
\sum_{n=1}^\infty \frac{(-\varrho)^{n-1}}{n!} \times\\
\times \int\limits_{\Omega}\rmd\Gamma_n\ 
\frac{M_\nu\big(\bigcap_{i=1}^n B_r(\bx_i)\big)}{M_\nu(B_r) |\Omega|}\
\xi_n(\bx_1,\ldots,\bx_n)    
\end{multline}
describe the deviation from a Poisson process (compare 
Eq.~\eqref{eq:mean1}).  For Poisson distributed spheres with vanishing 
cumulants $\xi_n(\Gamma_n)=0$ one recovers $\specM_\nu=1$.  The 
specific Minkowski functionals depend on the radius $r$, the product 
density $\varrho$, and all of the correlation functions 
$\xi_n(\bx_1,\ldots,\bx_n)$.  In particular, one obtains for the 
normalized Minkowski functionals in three dimensions 
($\eta=\frac{4\pi}{3}r^3 \varrho$)
\begin{equation} 
\label{eq:meancorr4}
\begin{array}{l}
\Phi_0(r;\varrho,\{\xi_n\}) = \Big(1-\rme^{-\eta \specM_{0}}\Big)/\eta  \\
\Phi_1(r;\varrho,\{\xi_n\})  = \specM_{1}\ \rme^{-\eta \specM_{0}} \\
\Phi_2(r;\varrho,\{\xi_n\}) =\Big(\specM_{2}-\frac{3\pi^2}{32}
\eta \specM_{1}^2\Big)\ \rme^{-\eta \specM_{0} } \\  
\Phi_3(r;\varrho,\{\xi_n\}) =
\Big(\specM_{3}-3\eta \specM_{1}\specM_{2}+\frac{3\pi^2}{32}
\eta^2\specM_{1}^3\Big)\ \rme^{-\eta \specM_{0}} \; ,   
\end{array}
\end{equation}  
which can be compared to the analogous result~\eqref{eq:mean1} for 
Poisson distributed grains.  The dependence on higher--order 
correlation functions enters only into a finite number of relevant 
coefficients $\specM_\nu$.  The universal polynomial form of the mean 
values~\eqref{eq:meancorr4} is related to the 
additivity~\eqref{eq:additivity} of the Minkowski functionals due to 
the decomposition into {\em specific} terms $\specM_\nu$.
We also define the deviations
\begin{equation}
\delta\specM_\nu(r) = \big(1 - \specM_\nu(r)\big) \frac{2^{1-\nu}}{\eta } 
\end{equation}
from  the  uncorrelated  Poisson values  $\specM_\nu^{(P)}(r)=1$  with
$\delta\specM_\nu^{(P)}(r)=0$.

We tried to use the {\em specific} Minkowski functionals $\specM_\nu$
or $\delta\specM_\nu$ to compare the cluster distribution with our
models.  However, due to the nonlinear dependence of the $\specM_\nu$
on the measured $\Phi_\nu$, the relative errors are significantly
enlarged, compared to the errors of the $\Phi_\nu$.  The
discriminatory power of the $\Phi_\nu$'s is lost.  This may be
understood in detail by solving Eq.~\eqref{eq:meancorr4} for the
$\specM_\nu$ and inserting an error
$\tilde{\Phi}_\nu(r)=\Phi_\nu(r)+\Delta\Phi_\nu(r)$ for the measured
values of the Minkowski functionals $\Phi_\nu$.  Expanding in powers
of $\Delta\Phi_\nu$ one gets
$\Delta\specM_\nu\propto\rme^{-\eta\specM_0}$ for any $\specM_\nu$.
The errors of the specific functionals increase exponentially for
large $\eta$.  It is clear that the $\Phi_\nu$ are the preferable
choice in the comparison of data with the models, but for the
analytical calculations the $\specM_\nu$ and $\delta\specM_\nu$ are
more appropriate.

%%%
\subsection{The Gauss--Poisson process}
\label{sect:gauss-poisson}

The notion of a Gaussian random field is well understood in cosmology
(e.g.\ {}\citealt{bardeen:gauss}): considering the density contrast
$\delta(\bx)=\rho(\bx)/\rho_H-1$ the two--point correlation function
$\xi^\delta_2(r)=\BE[\delta(0)\delta(r)]$ together with the mean mass
density $\rho_H$ specifies the statistical properties of the mass
density field completely ($\BE$ is the average over several
realizations of the random field).  The higher correlation functions
$\xi^\delta_n$ all equal zero.
In the following we will show how to construct a ``Gaussian'' point 
distribution.  A detailed discussion, examples, and extensions to 
higher--order processes is presented in 
{}\citet{kerscher:constructing}.  The defining property of this 
Gauss--Poisson point process, similar to the Gaussian random field, is 
that the higher--order correlation functions of the point set vanish: 
$\xi_n=0$ for $n>2$.  Due to the discrete nature, and the demand for a 
positive number density $\varrho$, some constraints on the two--point 
correlation function $\xi_2$ as well as the number density $\varrho$ 
emerge. 

In general, a point process may be specified by its probability
generating functional (p.g.fl.)  $G[h]$ where $h(\cdot)$ are suitable
functions (see e.g.\ {}\citealt{daley:introduction} Sect.~7.4;
$G[h]=\CR[h]$ as defined by {}\citealt{balian:I}).  The p.g.fl.\ is
the point process analogue of the probability generating function of a
discrete random variable {}\citep{kendall:advanced1}.  The expansion
of $G[h]$ in terms of the (connected) correlation functions $\xi_n$
reads:
\begin{multline}
\label{eq:pgfl-xin}
\log G[h+1] 
= \sum_{n=1}^\infty    
\frac{\varrho^n}{n!}\times \\
\int_{\BR^d}\rmd\bx_1\cdots\int_{\BR^d}\rmd\bx_n \
\xi_n(\bx_1,\ldots,\bx_n)\ h(\bx_1)\cdots h(\bx_n).
\end{multline}
If we truncate this expansion after $n=1$ we arrive at the p.g.fl.\ of
the Poisson  process.  A truncation  after $n=2$, i.e.\  $\xi_n=0$ for
$n>2$, defines the Gauss--Poisson point process:
\begin{multline}
\label{eq:G-gauss}
\log G[h+1] = \varrho\int_{\BR^d}\rmd\bx_1\ h(\bx_1) + \\
+\frac{\varrho^2}{2}\int_{\BR^d}\rmd\bx_1\int_{\BR^d}\rmd\bx_2\ 
\xi_2(|\bx_1-\bx_2|)\ h(\bx_1)h(\bx_2).
\end{multline}
The Gauss--Poisson point process ({}\citealt{newman:new},
{}\citealt{milne:further}) is stochastically fully specified by its
two--point correlation function $\xi_2(r)$ and the number density
$\varrho$. However, $\varrho$ and $\xi_2(r)$ may not be chosen
arbitrarily.  
{}\citet{milne:further} showed that the Gauss--Poisson process is only
well--defined if two constraints are satisfied. In
Appendix~\ref{sect:derive-constraints} we give a detailed derivation.

A simplified version of the constraint {}\eqref{eq:const1} is
\begin{equation}
\label{eq:const2-alternate}
\varrho\ \int_{A}\rmd\by\ \xi_2(|\by|) \le 1.
\end{equation}
This tells us that sitting on a point of the process on average at
most one other point in excess of Poisson distributed points is
allowed.
The constraint~\eqref{eq:const2} implies 
\begin{equation}
\label{eq:const1-alternate}
\xi_2(r)\ge0\ \text{ for any }\ r,
\end{equation}
hence only clustering point distributions may be modeled as a
Gauss--Poisson process.

Clearly the question arises, what is wrong with the simple picture
that we start with a Gaussian random field and ``Poisson sample'' it
to obtain the desired point distribution.  The answer is that a
Gaussian random field is an approximate model for a mass density field
only if the fluctuations are significantly smaller than the mean mass
density.  Otherwise negative mass densities (i.e.\ negative
``probabilities'' for the Poisson sampling) would occur. Only in the
limit of vanishing fluctuations a Poisson sampled Gaussian random
field becomes a permissible model. However, in this limit we are left
with a pure Poisson process.

%%%%
\subsection{Simulating the Gauss--Poisson process}
\label{sect:simulate-gauss-poisson}

As   discussed  by  {}\citet{daley:introduction}   any  Gauss--Poisson
process  {\em  equals}  a   rather  simple  type  of  Poisson  cluster
processes (for details see {}\citealt{kerscher:constructing}).
A Poisson cluster processes is a two--stage point process.  First we
distribute parent points $\by$ (the supercluster centers) according
to a Poisson process with number density $\varrho_p$ and then we
attach to each parent a second point process (the supercluster).  In
this specific example the supercluster consists only of one or two
points with probability $q_1(\by)$ and $q_2(\by)$, respectively.  We
have $q_1(\by)+q_2(\by)=1$ and the first point is the supercluster
center itself.  The probability density $f(|\bx-\by|)$ determines the
distribution of the distance $|\bx-\by|$ of the second point $\bx$ to
the supercluster center with $\int\rmd\bx f(|\bx|)=1$.
The p.g.fl.\ of this Poisson cluster process is given by
\begin{multline}
\log G[h+1] = \int_{\BR^d}\rmd\bx_1\ \varrho_p(1+q_2(\bx_1))\ h(\bx_1) +\\
+ \int_{\BR^d}\rmd\bx_1\int_{\BR^d}\rmd\bx_2\ 
\varrho_p q_2(\bx_1) f(|\bx_1-\bx_2|)\ h(\bx_1)h(\bx_2),
\end{multline}
which equals the p.g.fl.~\eqref{eq:G-gauss} for the Gauss--Poisson
process for $\varrho=\varrho_p(1+q_2)$ and
$\xi_2(r)=2\frac{\varrho-\varrho_p}{\varrho^2}f(r)$
({}\citealt{daley:introduction} Sect.~8.3,
{}\citealt{kerscher:constructing}).  Hence, {\em every} Gauss--Poisson
process is a Poisson cluster process of the above type, and vice
versa.
For a Gauss--Poisson process it is necessary that the parent points,
in our case the supercluster centers, are distributed according to a
Poisson process.  Any deviation from a pure Poisson process, either in
the direction of clustering or regularity, implies the presence of
higher--order correlation functions in the distribution of the galaxy
clusters.  There are indications that the distribution of galaxies or
galaxy clusters shows some regularity on large scales
{}\citep{broadhurst:large-scale,einasto:120mpc}. Hence, an unambiguous
detection of such large scale regularity in the upcoming large
redshift surveys would also strengthen our findings of non--Gaussian
features on large--scale (see below). However, regular features on
large scales are not necessary for higher--order correlations to be
present.

We may generate realizations of the Gauss--Poisson process for a given
number density $\varrho$ and two-point correlation function
$\xi_2(r)$, fulfilling the constraints~\eqref{eq:const1} and
{}\eqref{eq:const2}.  With $C_2=\int_{\BR^d}\rmd\bx\ \xi_2(|\bx|)$ and
$\int_{\BR^d}\rmd\bx\ f(|\bx|)=1$ we can calculate the quantities
needed in the simulation: $f(r)=\xi_2(r)/C_2$, $q_2=\frac{\varrho
C_2}{2-\varrho C_2}$, $q_1=1-q_2$, and $\varrho_p=\varrho(1-\varrho
C_2/2)$.  The constraint~\eqref{eq:const2-alternate} implies
$C_2\varrho\le1$.  The simulation is carried out in two steps:
\begin{itemize}
\item First we generate the parents (the supercluster centers)
according to a Poisson distribution with number density $\varrho_p$.
\item For each supercluster center $\by$ we draw a uniform random 
number $q$ in $[0,1]$.  If $q<q_1$, then we keep only the point $\by$.  
If $q\ge q_1$ then additionally to the point $\by$ we  chose a 
random direction on the unit sphere and a distance $d$ with the 
probability density $f$ and place the second point according to them.
\end{itemize}
To get the correct point pattern inside the sample window, one also
has to use supercluster centers outside the window to make sure that
any possible secondary point is included in the sample inside the
window.

\begin{figure}
\begin{center}
\epsfig{file=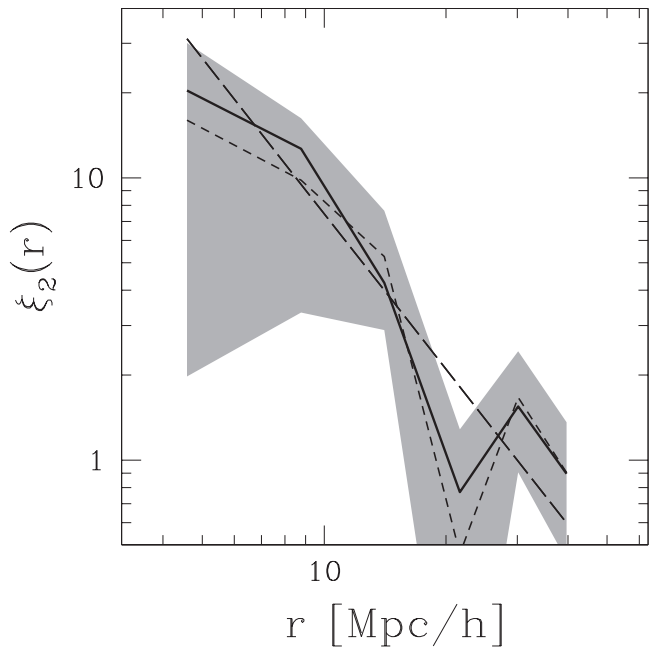,width=9cm}
\end{center}
\caption{\label{fig:xiL20} The two--point correlation function 
$\xi_2(r)$ estimated from the L20 sample (solid line) is shown 
together with $\xi_2(r)$ estimated from 200 realizations of a 
Gauss--Poisson process with the same $\xi_2$ and number density as for 
the observed clusters (short dashed line, shaded one--$\sigma$ area).  
The long dashed line is for $\xi_2(r)=(30/r)^{1.83}$.}
\end{figure}

As an illustration of this procedure we calculate the two--point
correlation function $\xi_2$ from the sample L20. The $\xi_2$ together
with the number density satisfy the
constraint~\eqref{eq:const2-alternate} (see
Table~\ref{table:samples}): $\varrho C_2<1$.  We use this empirical
$\xi_2$ as an input to the simulation algorithm outlined above.
Fig.~\ref{fig:xiL20} illustrates that these simulated Gauss--Poisson
sets are indeed able to reproduce the observed two--point correlation
function.  Even the dip of $\xi_2$ at 20\hMpc, is well reproduced by
the simulated point sets.  By construction no higher--order
correlations are present in the simulated point sets in the mean.

\subsection{Minkowski functionals of a Gauss--Poisson process}
\label{sect:minkowski-gauss-poisson}

In the following the Minkowski functionals for a Gauss--Poisson
process will be given.  Truncating after the second term in
Eq.~\eqref{eq:meancorr3} one obtains the correlated average of the
specific Minkowski functionals
\begin{align} 
\label{eq:minkgeneral2}
\specM_\nu(r;\varrho,\xi_2) = & 1 - 2^{\nu-1}\eta\ 
\delta\specM_\nu^{(GP)}(r;\xi_2), \\[1ex]
\delta \specM_\nu^{(GP)}(r;\xi_2) = & \frac{1}{2^{\nu}M_0(B_r)}
\int\limits_{B_{2r}(\mathbf{0})}\rmd\bx\ \xi_2(|\bx|)\ I_\nu(r;|\bx|)
\nonumber   
\end{align} 
with the volume $M_0(B_r)$ of a sphere of radius $r$.  
From Eq.~\eqref{eq:minkgeneral2} and using the expressions for
$I_\nu(r;s)$ given in Eq.~\eqref{eq:define-Imu-3d} we can calculate
the normalized Minkowski functionals of the Gauss--Poisson process
according to Eq.~\eqref{eq:meancorr4}.  For example we obtain
\[
\Phi_0(r;\varrho,\xi_2) = \big(1-\rme^{-\eta \specM_{0}}\big)/\eta,
\]
with
\[
\specM_0 = 1-\frac{\eta\ 4\pi}{2M_0(B_r)}
\int_0^{2r}\rmd s\ s^2 \xi_2(s) \Big( 1 - \frac{3}{2} \frac{s}{2r}  + 
\frac{1}{2}\big(\frac{s}{2r}\big)^3 \Big),
\]
and similar expressions for $\Phi_1$, $\Phi_2$, and $\Phi_3$.

The expansion {}\eqref{eq:eta-expansion} to linear order in
$\eta\propto r^3\varrho$ allows us to describe the MFs only for small
radii or low number densities.  For a Gauss--Poisson process both
$I_\mu$ and $\eta$ appear non--linearly in Eq.~\eqref{eq:meancorr4}
via Eqs.~(\ref{eq:corradd}, {}\ref{eq:minkgeneral2},
{}\ref{eq:minkgeneral2}).  Contrary to the approximation
{}\eqref{eq:eta-expansion} which is only valid for $\eta\ll1$, the
Minkowski functionals of the Gauss--Poisson process, are valid for all
$\eta$.  A Gauss--Poisson process does not imply the linearity of the
MFs in $I_\mu$ and $\eta$.

For a Poisson distribution with  $\xi_2(r)=0$ one obtains
\begin{equation} 
\specM_\nu^{(P)}(r) = 1 \;, \;\; \delta \specM_\nu^{(P)}(r) = 0 \;, 
\end{equation}
i.e., one recovers  Eq.~\eqref{eq:mean3d}.
Assuming an algebraic  scaling form 
\begin{equation}
\label{eq:scalingxi}
\xi_2(s) = \Big(\frac{s_0}{s}\Big)^\gamma 
\end{equation}
for the  centered two--point correlation function one obtains with
$\gamma<3$ the general result 
\begin{equation} 
\label{eq:scalingmnu}
\delta \specM_\nu^{(GP)}(r) =  
A_\nu(\gamma)\ \Big(\frac{s_0}{2r}\Big)^\gamma   
\end{equation}
with the amplitudes 
\begin{equation} 
\label{eq:amplitudes}
\begin{array}{ll}
A_0(\gamma) = & \frac{24}{3-\gamma}-\frac{36}{4-\gamma}+\frac{12}{6-\gamma},\\ 
A_1(\gamma) = & \frac{12}{3-\gamma} -  \frac{12}{4-\gamma} ,\\ 
A_2(\gamma) = & \frac{6}{3-\gamma} -  \frac{6}{4-\gamma} +
 3\int\limits_0^{\pi/2} \rmd y\ y \cos^2y \sin^{2-\gamma}y ,\\ 
A_3(\gamma) = & \frac{3}{3-\gamma} . \\ 
\end{array} 
\end{equation}
In the limiting case $\gamma\rightarrow0$ (with $\varrho\rightarrow0$)
one recovers the result $A_0=A_1=A_3=1$ and $A_2=\frac{1}{2} +
\frac{3\pi^2}{64}$ for the averaged Minkowski functionals
\begin{equation} 
\label{eq:amplitudes2}
A_\nu \; = \; 2^{-\nu} \int \rmd\bx \; 
\frac{M_\nu(B_r(\mathbf{0})\cap B_r(\bx))}{M_\nu(B_r)M_0(B_r)  }
\end{equation} 
of  the   intersection  of  two  overlapping   spheres  with  distance
$|\bx|$ and radius $r$ (see kinematic formula~\eqref{eq:kinematic}).

One has to be careful in the interpretation of
Eq.~\eqref{eq:amplitudes}, since a scale--invariant two--point
correlation function~\eqref{eq:scalingxi} with $\gamma<3$ does not
satisfy the constraint~\eqref{eq:const2-alternate}, as required for
the existence of a Gauss--Poisson process.  A cut--off, i.e.\
$\xi_2(s)=0$ for $s>r_c$, has to be imposed below some radius $r_c$ to
guarantee the constraint~\eqref{eq:const2-alternate}.  The maximal
allowed $r_c$ is depending on the number density
through~\eqref{eq:const2-alternate}.  As can be seen directly from
Eq.~\eqref{eq:minkgeneral2}, the scaling
behavior~\eqref{eq:scalingmnu} as well as the amplitudes $A_\nu$ in
Eq.~\eqref{eq:amplitudes} are still correct for $2r<r_c$.  Only on
larger scales additional terms depending on the cut--off will emerge.
In such a scale invariant Gauss--Poisson process the specific MFs
$\delta\specM_\nu(r)$ should show the general scaling
form~\eqref{eq:scalingmnu} which may be used to test for an algebraic
two--point correlation function $\xi_2(r)$.

The actual measured specific MFs still depend on the density $\varrho$
and all of the correlation functions
$\xi^{(n)}(\bx_1,\ldots,\bx_n)$. The deviations of the measured MFs
from the expressions~\eqref{eq:minkgeneral2} for a Gauss--Poisson
process will be used as a measure for the relevance of higher--order
correlations among the points (see Sect.~\ref{sect:gauss-cluster}).

%%%%%

To  facilitate future applications  we will  also quote  the Minkowski
functionals   of   a   Gauss--Poisson   process   in   two   dimensions
{}\citep{mecke:diss}. With $\eta=\pi r^2\varrho$ the reduced Minkowski
functionals (compare Eq.~\eqref{eq:meancorr4} for three dimensions)
\begin{equation}
\label{meancorr2d}
\begin{array}{l}
\Phi_0(r;\varrho,\{\xi_n\}) = (1-\rme^{-\eta \specM_{0}})/\eta  \\
\Phi_1(r;\varrho,\{\xi_n\})  = \specM_{1}\rme^{-\eta \specM_{0}} \\
\Phi_2(r;\varrho,\{\xi_n\}) =(\specM_{2}-\eta \specM_{1}^2)\rme^{-\eta \specM_{0} } ,   
\end{array}
\end{equation}  
with the specific Minkowski functionals given by
Eq.~\eqref{eq:meancorr3}.  For a Gauss--Poisson process the specific
Minkowski functionals are given by the Eq.~\eqref{eq:minkgeneral2},
the volume $M_0(B_r)=\pi r^2$ of a disc of radius $r$, and the
functions
\begin{equation} \label{area2}
\begin{array}{rl} 
I_0(r;s) = &  1 - \frac{2}{\pi} 
\bigg(\arcsin\frac{s}{2r} - \frac{s}{2r} 
% \sqrt{1-\left(\frac{s}{2r}\right)^2}
 \Big(1-\big(\frac{s}{2r}\big)^2\Big)^{\tfrac{1}{2}}
\bigg) \\ 
I_1(r;s) = & 1 - \frac{2}{\pi} \arcsin\frac{s}{2r}  \\ 
I_2(r;s) = & \Theta(2r-s) . \\ 
\end{array} 
\end{equation}  
If      the      correlation      function     decays      algebraicly
$\xi(r)=(s_0/r)^\gamma$, $r=|\bx_1-\bx_2|$, with an scaling
exponent $\gamma$  and a  correlation length $s_0$  one finds  for the
intensities   of  the   Minkowski   measures~\eqref{eq:meancorr4}  for
homogeneously distributed discs $B_r$  of radius $r$ in two dimensions
the amplitudes {}\citep{mecke:diss}
\begin{equation} \label{eq:amplitudes2d}
\begin{array}{ll}
A_0(\gamma)= & \frac{8}{2-\gamma }-2g(\gamma)-f(\gamma) , \\
A_1(\gamma)  = & \frac{4}{2-\gamma}-g(\gamma) , \\  
A_2(\gamma)  = & \frac{2}{2-\gamma} ,
\end{array} 
\end{equation}
with the functions
\begin{equation}
f(\gamma)=
%\frac{16}{\pi} \int\limits_0^1\rmd y y^{2-\gamma}\sqrt{1-y^2} =
\frac{4}{\sqrt{\pi}} 
\frac{\Gamma\big(\tfrac{3}{2}-\tfrac{\gamma}{2}\big)}{\Gamma\big(3-\tfrac{\gamma}{2}\big)} ,
\end{equation}
and 
\begin{align}
g(\gamma)
& = \frac{8}{\pi }\int\limits_0^1\rmd y\ y^{1-\gamma} \arcsin(y) .
%\nonumber\\
%& = \frac{8}{\pi}\int\limits_0^{\pi/2}\rmd y\ y\ \cos y\ \sin^{1-\gamma}y . 
\end{align} 
One also recovers the result $A_\nu=1$ in the limit
$\gamma\rightarrow0$ (with $\varrho\rightarrow0$) for the averaged
Minkowski functionals of the intersection of two discs defined in
Eq.~\eqref{eq:amplitudes2}.
In general a cut--off in the scale invariant correlation function is
needed to make this model well--defined. See the comments above.

\section{Non--Gaussian morphology of the galaxy cluster distribution}
\label{sect:gauss-cluster}

In this section we compare the Minkowski functionals determined from
the cluster distribution with the Minkowski functionals of a
Gauss--Poisson process.  In Sect.~\ref{sect:gauss-poisson} we showed
that the number density $\varrho$ and the two--point correlation
function $\xi_2(r)$ have to fulfill constraints in order to allow them
to serve as the ingredients for a Gauss--Poisson point process.
The constraint {}\eqref{eq:const2} implies $\xi_2(r)\ge0$ for all $r$.
There are indications from the analysis of the flux--limited {}\reflex
catalogue that the two--point correlation function of the cluster
distribution becomes negative on scales of 40--50\hMpc\
{}\citep{collins:spatial}.  The violation of constraint
{}\eqref{eq:const2} already tells us that the cluster distribution
exhibits non--Gaussian features even on such large scales.  Due to the
limited number of clusters and the smaller extent of samples we can
not detect this zero crossing unambiguously in the volume--limited
samples we analyzed.  To obtain a well--defined model of the two--point
correlation function we impose a cut at $r_c=48\hMpc$ with
$\xi_2(r)=0$ at $r>r_c$.

Another more stringent constraint is Eq.~\eqref{eq:const1} which may be 
cast into the form
\begin{equation}
\label{eq:const1-ii}
\varrho\ \int_{\BR^3}\rmd\by\ \xi_2(|\by|) = \varrho\ C_2 \le 1.
\end{equation}
As can be seen from Table~\ref{table:samples}, the smaller volume--limited samples L05--L12, with their higher number density clearly
violate this constraint.  Hence, already by inspecting the two--point
correlation function $\xi_2$ in conjunction with the number density
$\varrho$ we can conclude that there are non--negligible higher--order
correlations in the point distribution of galaxy cluster in the
samples L05 and L12.

In Sect~\ref{sect:simulate-gauss-poisson} we showed how to simulate a
Gauss--Poisson process for a given two--point correlation function
$\xi_2$ and $\varrho$.  The violation of the constraint
{}\eqref{eq:const1-ii} prohibits the simulation of a Gauss--Poisson
process corresponding to the samples L05 and L12.
However for the samples L20 and L30 the constraint
{}\eqref{eq:const1-ii} is satisfied (see Table~\ref{table:samples})
and we may generate realizations of a Gauss--Poisson process with the
same number density and the same two--point correlation function, as
estimated from these samples, (see Sect.~\ref{sect:gauss-poisson} and
especially Fig.~\ref{fig:xiL20}).  Both constraints
{}\eqref{eq:const2-alternate} and {}\eqref{eq:const1-alternate} are
only necessary conditions for the existence of a Gauss--Poisson
process. Still higher--order correlations may be present in the
cluster distribution.  We calculate the Minkowski functionals of these
Gauss--Poisson samples and compare them with the Minkowski functionals
of the observed cluster distribution.  Since correlations of any order
enter the Minkowski functionals (see Eq.~\eqref{eq:corradd} or
Eqs.~\eqref{eq:meancorr3} and {}\eqref{eq:meancorr4}), deviations of
the Minkowski functionals of the cluster distribution from the
Minkowski functionals of the Gauss--Poisson process indicate the
presence of higher--order correlations even in these deep samples.
To  facilitate   the  comparison  we  use   the  normalized  Minkowski
functionals
\begin{equation}
\label{eq:phi-varrho}
\Phi_\nu(r) = \frac{\densM_\nu(\CA_r)}{M_\nu(B_r)\; \varrho(r)} .
\end{equation}
To get unbiased estimates of the MFs we employ a boundary corrected
estimator where we shrink the observational window $W(r)$ with
increasing radius $r$ of the spheres ({}\citealt{mecke:robust},
{}\citealt{schmalzing:cfa2}).  As an estimate of the number density
inside $W(r)$ we use $\varrho(r)=N(r)/|W(r)|$ with $N(r)$ the number
of points inside the shrunken window with the volume $|W(r)|$.

\begin{figure*}
\begin{center}
\epsfig{file=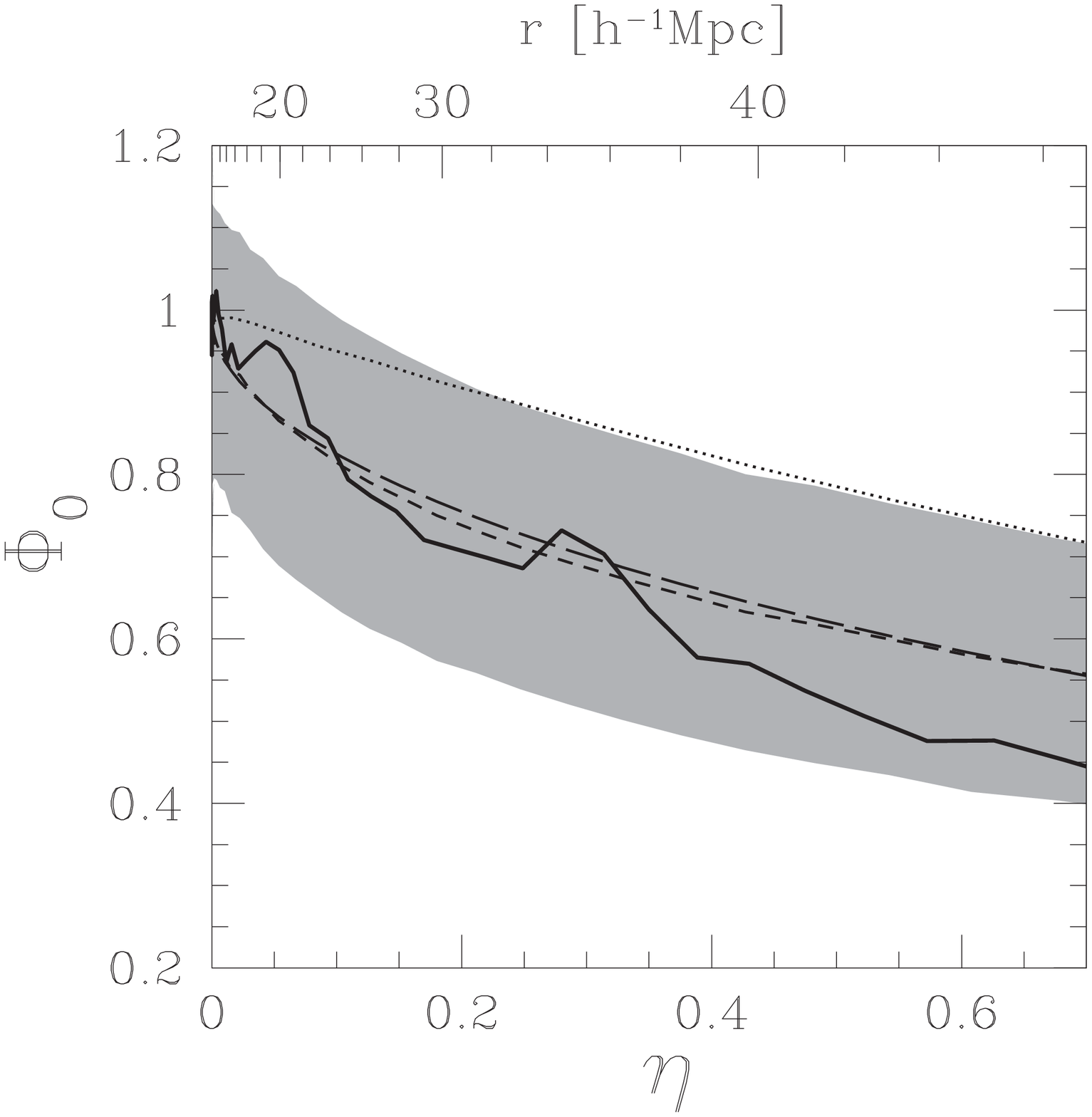,width=7.5cm}
\epsfig{file=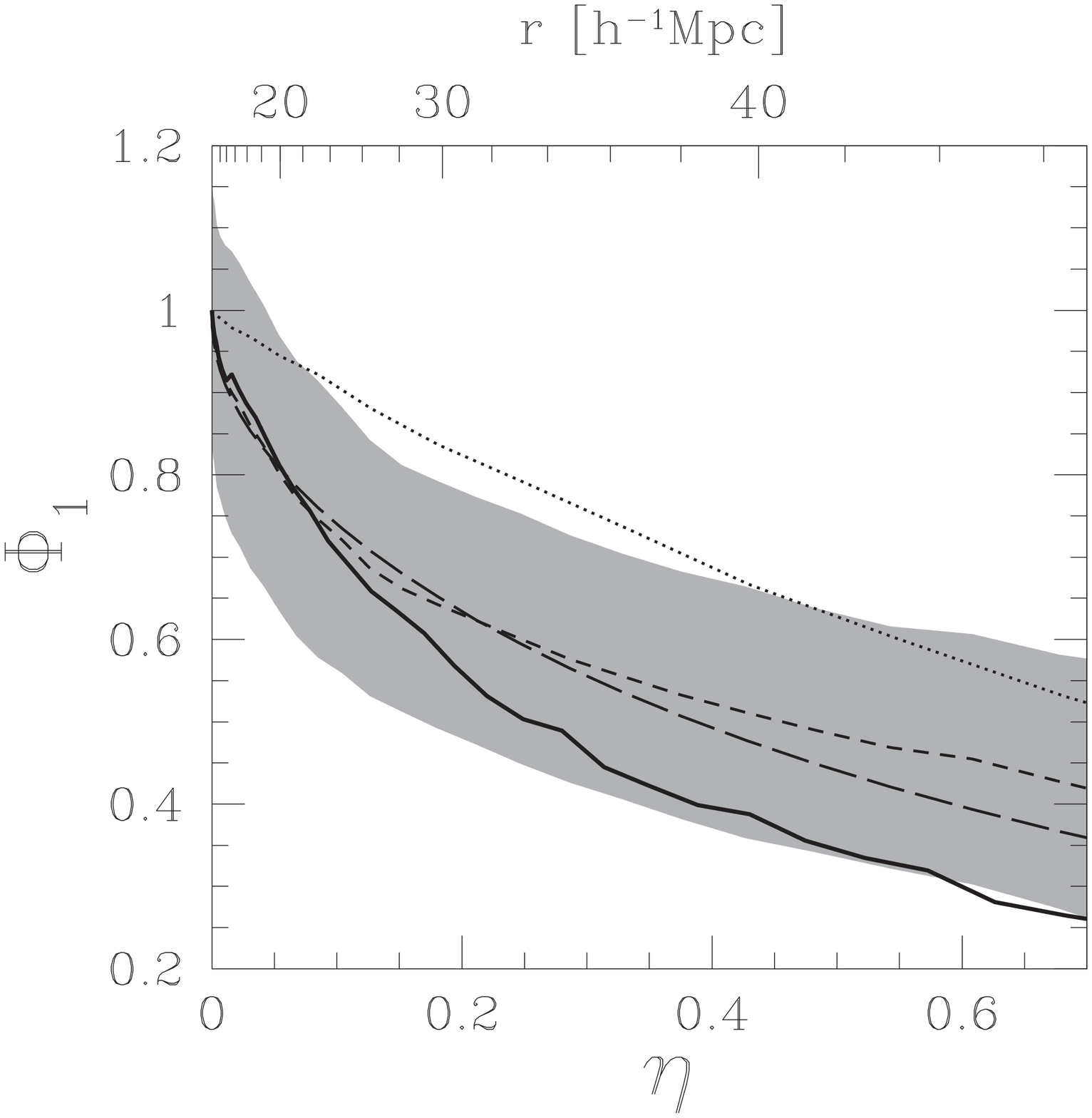,width=7.5cm}
\epsfig{file=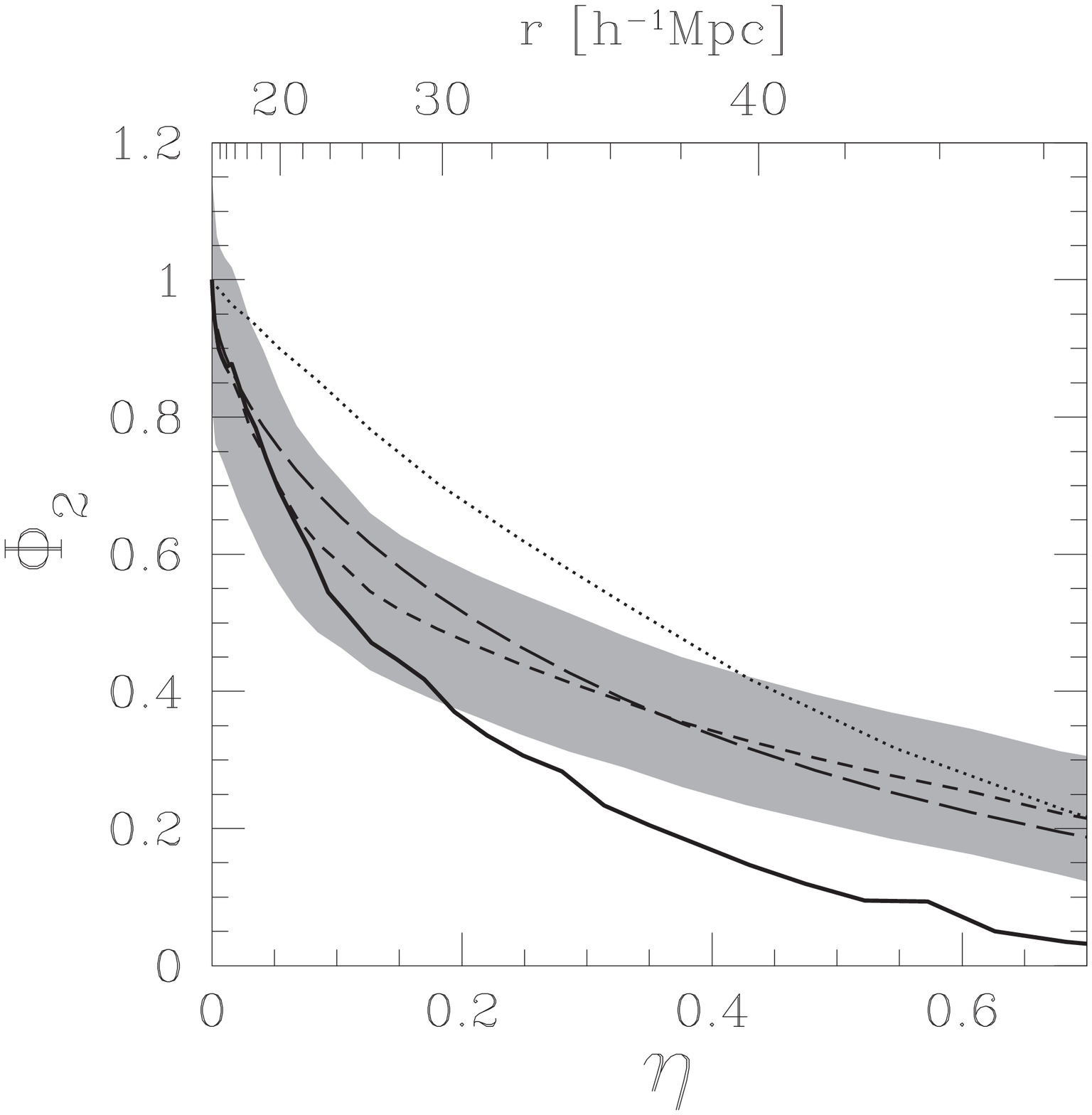,width=7.5cm}
\epsfig{file=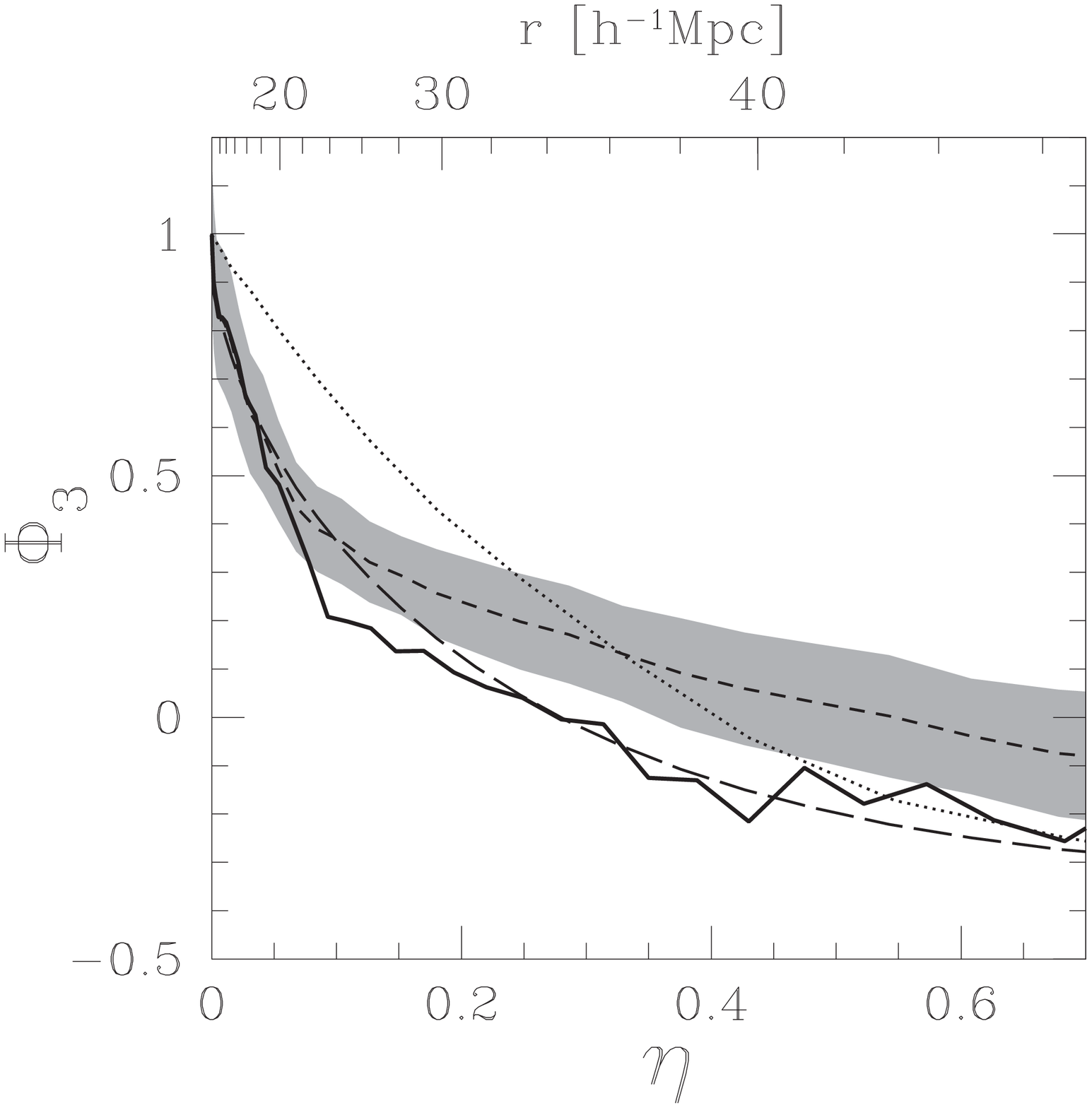,width=7.5cm}
\end{center}
\caption{\label{fig:PhigpL20} The reduced Minkowski functionals
$\Phi_\mu$ of the volume--limited sample L20 (solid line, see also
Fig.~\ref{fig:min20}) compared to the Minkowski functionals of a
Gauss--Poisson process (short dashed line, shaded one--$\sigma$ area)
with the empirical two--point correlation function as input (see
Fig.~\ref{fig:xiL20}).  The long dashed line is obtained from the
scale invariant model (see Eq.~\eqref{eq:scalingmnu}) with $s_0=30\hMpc$,
$\gamma=1.83$ and the amplitudes $A_0=6.8$, $A_1=4.73$, $A_2=2.70$,
and $A_3=2.56$ according to Eq.~\eqref{eq:amplitudes}.
The mean Minkowski functionals of a Poisson process are shown as a
dotted line.  The errors for the Poisson process (not shown) are
smaller than the errors from the Gauss--Poisson process.}
\end{figure*}

In Fig.~\ref{fig:PhigpL20} the results of our comparison are shown, where
the $\Phi_\mu$'s are plotted against $\eta=\varrho(4\pi/3)r^3$.  We used
the empirical two--point correlation function to generate the
realizations of the Gauss--Poisson process inside the sample geometry
of the {}\reflex cluster catalog.  The shaded one--$\sigma$ area with
the short dashed line in the center was estimated from one hundred
realizations.
The initial slope of the MFs of the cluster distribution (solid line)
is well approximated by the expression {}\eqref{eq:eta-expansion}. But
already for fairly small $\eta$ this linear approximation breaks
down. As discussed in Sect.~\ref{sect:minkowski-gauss-poisson} a
Gauss--Poisson process does not imply the linearity of the MFs in
$\eta$, which is readily observed in Fig.~\ref{fig:PhigpL20}. It is
necessary to compare the measured values of $M_\nu(r)$ with
Eq.~\eqref{eq:meancorr4} which is available in the analytic form
only because of additivity. So our heuristic argument in the beginning
(to look for additive integral information on higher correlations)
turns out to be useful in deriving analytic results which are
necessary for the comparison with measured values.
Over the whole range of scales probed, the normalized volume $\Phi_0$
and surface area $\Phi_1$ of the cluster samples are consistent with
the Gauss--Poisson process.  However, both the normalized integral
mean curvature $\Phi_2$ and the normalized Euler characteristic
$\Phi_3$ are lowered with respect to the Gauss--Poisson process,
clearly outside the one--$\sigma$ range.  
The deviations are especially prominent for radii larger than
30\hMpc. This is a firm indication that higher--order correlations are
necessary to account for the shape and topology of the cluster
distribution for such large scales, given by the radii of the spheres.

In Fig.~\ref{fig:PhigpL20} also the MFs of a Gauss--Poisson process
with a scale invariant two--point correlation function
$\xi_2(r)=(s_0/r)^\gamma$ are shown (see Eqs.~\eqref{eq:scalingmnu},
{}\eqref{eq:amplitudes} and {}\eqref{eq:meancorr4}).  
The exponent is
$\gamma=1.83$ as determined by {}\citet{collins:spatial} and
$s_0=30\hMpc$. These parameters give a reasonable fit to the wiggly
two--point correlation function determined from volume--limited sample
L20 (Fig.~\ref{fig:xiL20}).
The normalized volume $\Phi_0$, surface area $\Phi_1$ and integral
mean curvature $\Phi_2$ of a Gauss--Poisson process with this scale
invariant $\xi_2$ follow closely the corresponding quantities
determined from the Gauss--Poisson process using the two--point
correlation function from the data.  A significant difference between
the two Gaussian models shows up only in the Euler characteristic
$\Phi_3$.

To quantify the deviation of the distribution of galaxy clusters from
a Gauss--Poisson process we perform a non--parametric significance
test {}\citep{besag:simple,stoyan:basic}.
Using $M=10$ equidistant radii $r_i$ in the range from 30\hMpc\ to
45\hMpc\ we define the ``distance''
\begin{equation}
d_\nu^k = \frac{1}{M}\sum_{i=1}^M 
\Big(\Phi_\nu^k(r_i)-\Phi_\nu^{\text{GP}}(r_i)\Big)^2
\end{equation}
between the normalized Minkowski functionals $\Phi_\nu^k$ of the
sample labeled with $k$ and the mean value $\Phi_\nu^{\text{GP}}$ of
the Gauss--Poisson process calculated from one hundred realizations
inside the same sample geometry as the {}\reflex samples and using the
empirical $\varrho$ and $\xi_2$.  Additional to the distances
$d_\nu^{\text{L20}}$ of the cluster sample L20, we also calculate the
distances $d_\nu^k$ for $k=1\ldots99$ realizations of the
Gauss--Poisson process (these 99 realizations are independent from the
realizations used to calculate the mean $\Phi_\nu^{\text{GP}}$).  We
order these distances including $d_\nu^{\text{L20}}$ ascending.  If
$d_\nu^{\text{L20}}$ is under the five highest distance values, we may
exclude a Gauss--Poisson process with a significance of 95\% (see the
comments by {}\citealt{mariott:barnard} concerning the significance
level).  The beauty of this Monte--Carlo significance test is that we
neither make assumptions about the distribution of the {}\reflex
clusters, nor about the distribution of the errors of the MFs in the
model.

\begin{table}
\begin{center}
\caption{ \label{table:ranks} The rank of the cluster sample L20
within the ascending list of one hundred ordered distances
$d_\nu^k$. A rank larger than 95 indicates rejection with significance
of 95\%.}
\begin{tabular}{c|c|c|c|c}
$\nu$ & 0 & 1 & 2 & 3 \\
\hline
rank & 26 & 52 & 97 & 96
\end{tabular}
\end{center}
\end{table}
In Table~\ref{table:ranks} the rank of the cluster sample L20 within
the ordered list of distances is given.  As expected from the visual
impression in Fig.~\ref{fig:PhigpL20} the volume $\Phi_0$ and the
surface area $\Phi_1$ are consistent with the expectation from a
Gauss--Poisson process.
Both the integral mean curvature $\Phi_2$ and the Euler characteristic
$\Phi_3$ allows us to reject the hypotheses that the cluster
distribution stems from a Gauss--Poisson process at a significance
level of 95\%.  This result is stable against extending or shrinking
the radial range. One may also use the overall  mean number density
$\varrho$ of the full sample in Eq.~\eqref{eq:phi-varrho} instead of
$\varrho(r)$. 

We implemented this Monte--Carlo test using the empirical two--point
correlation function as input to the simulations of the Gauss--Poisson
process. For a scale--invariant $\xi_2(r)=(30\hMpc/r)^{1.83}$ the
normalized Euler characteristic $\Phi_3$ according to
Eq.~\eqref{eq:scalingmnu} seems to be in agreement with the observed
MFs.  However, the family of MFs provides us with a consistency
check, still the integral mean curvature $\Phi_2$ from the scale
invariant model and the data are differing, illustrating the relevance
of higher--order correlations. Hence, also a Gauss--Poisson process
with a scale--invariant correlation function is inconsistent with the
data.

\begin{figure*}
\begin{center}
\epsfig{file=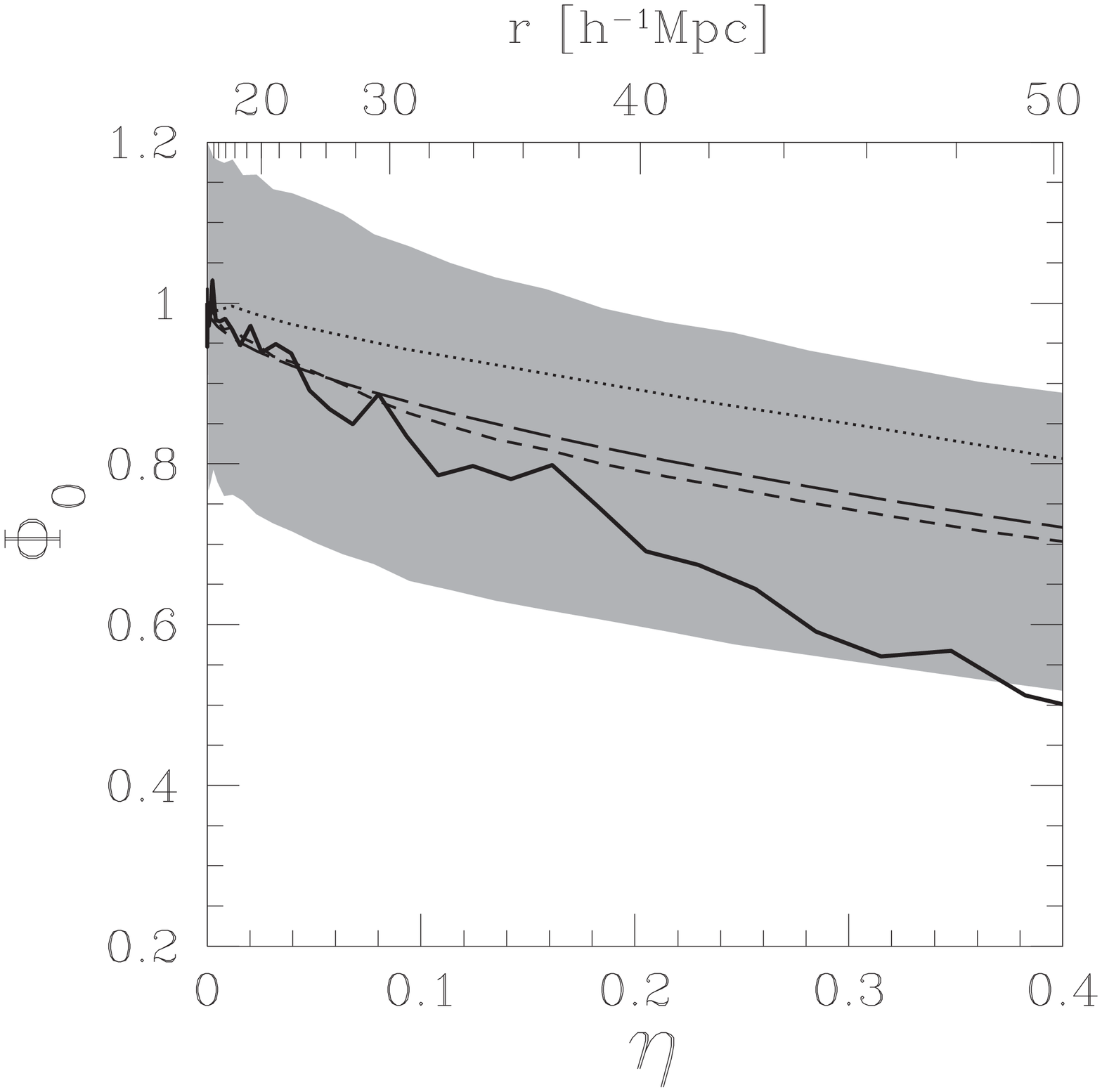,width=7.5cm}
\epsfig{file=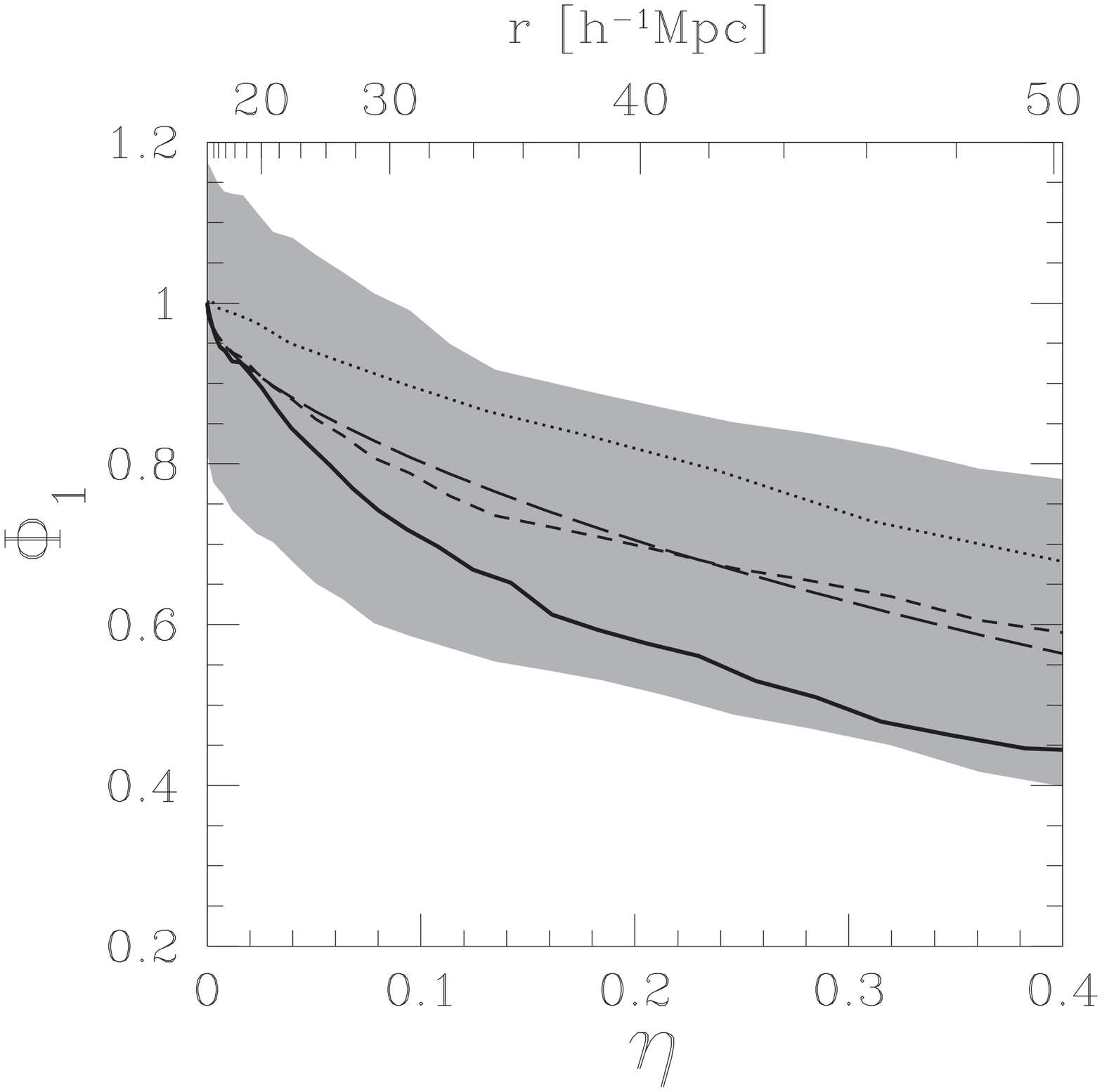,width=7.5cm}
\epsfig{file=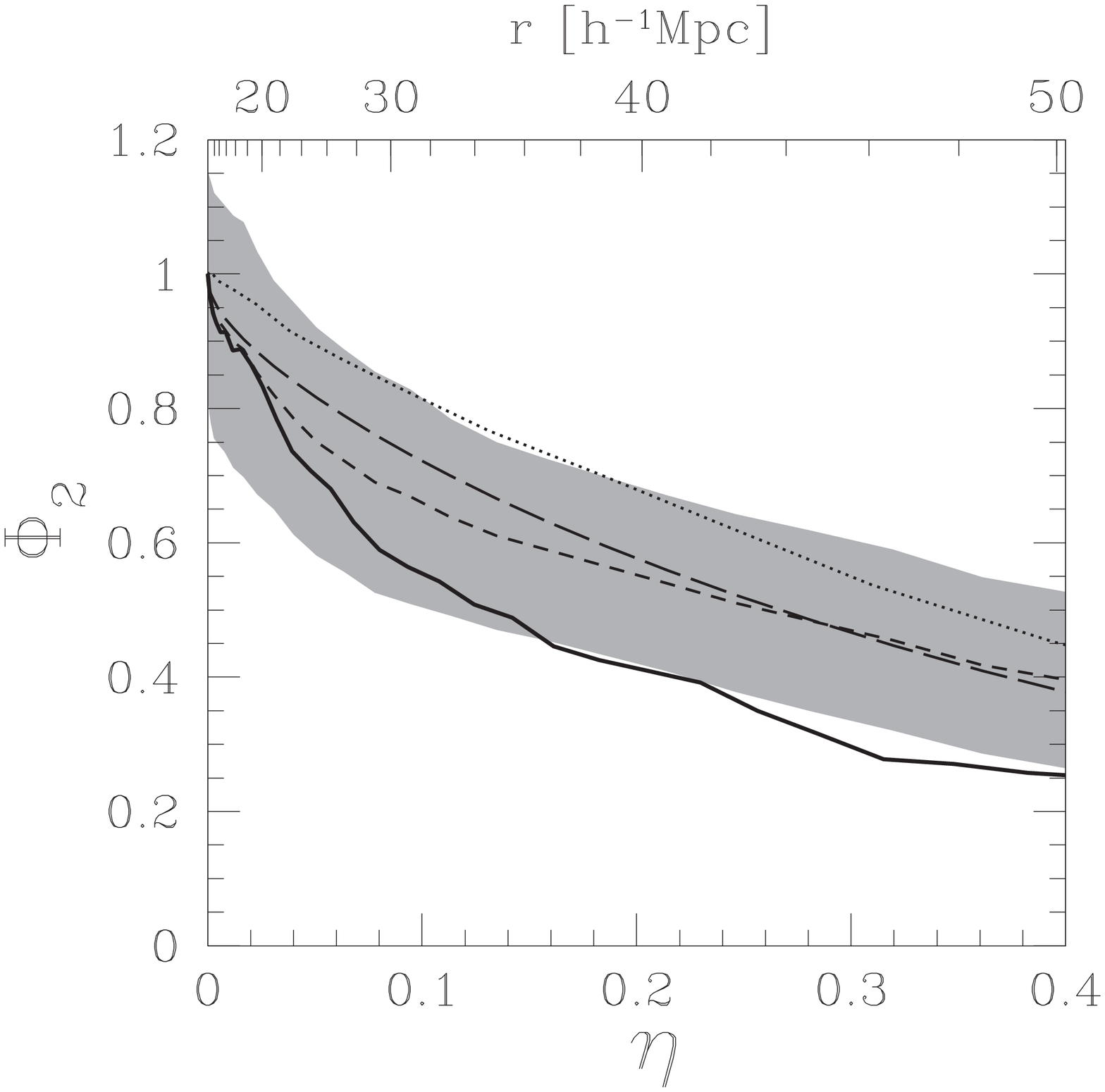,width=7.5cm}
\epsfig{file=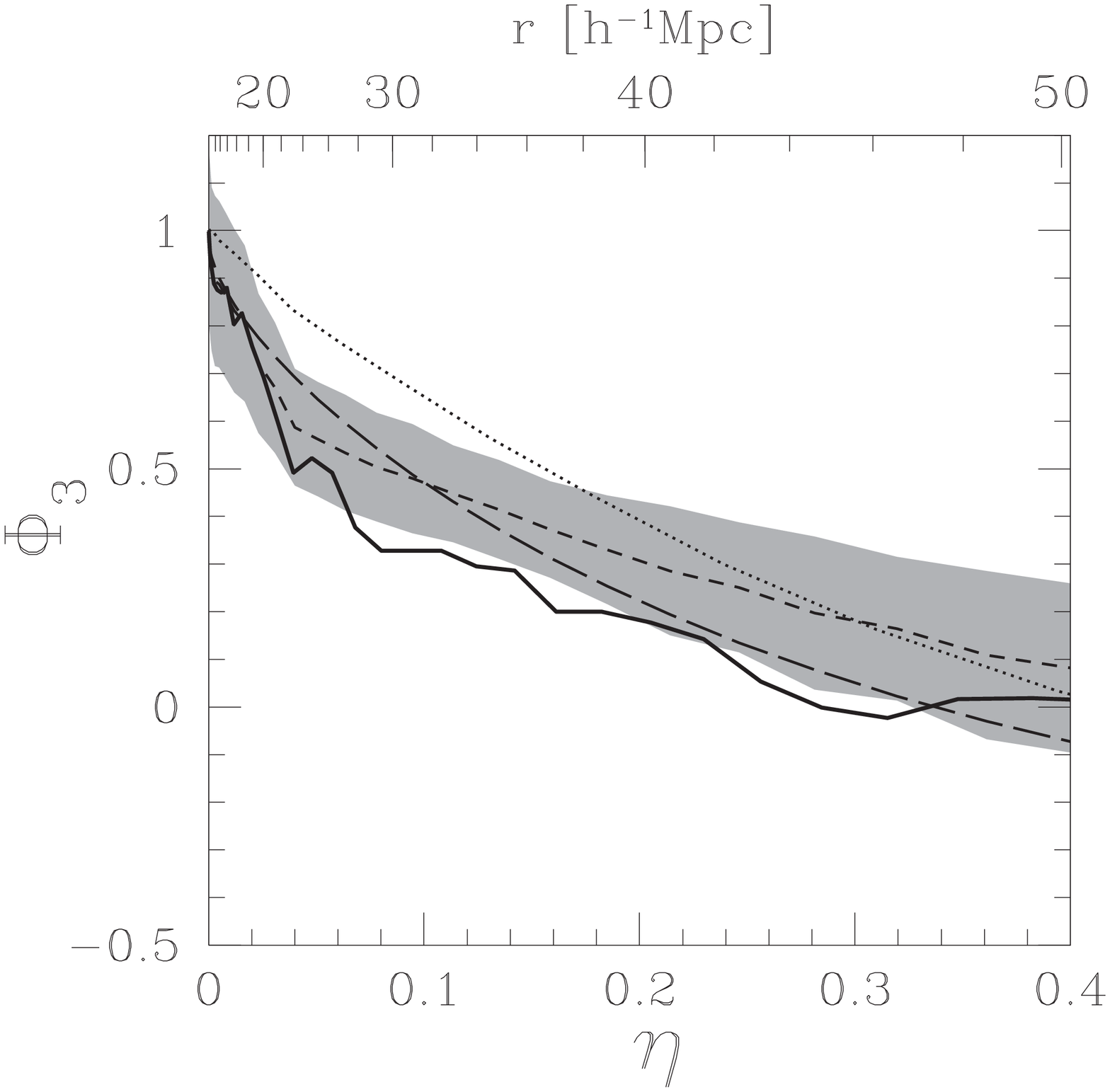,width=7.5cm}
\end{center}
\caption{\label{fig:PhigpL30} The reduced Minkowski functionals
$\Phi_\mu$ of the volume--limited sample L30 (see also
Fig.~\ref{fig:min30}).  The same conventions as in
Fig.~\ref{fig:PhigpL20} apply.}
\end{figure*}
We conducted a similar analysis for the cluster sample L30 (see
Fig.~\ref{fig:PhigpL30}).  Again $\Phi_0$ and $\Phi_1$ fall within the
one--$\sigma$ range of the Gauss--Poisson process. The $\Phi_2$ and
$\Phi_3$ only marginally stand out.  As discussed in
Sect.~\ref{sect:minkowski-reflex} this sample hardly allows for a
discrimination from the Poisson process, which can be explained by
Eq.~\eqref{eq:eta-expansion} and the significantly lowered number
density. Nevertheless the same tendency can be observed as for the
sample L20 although the statistics does not allow a discrimination.

%%%%
\section{Summary}

The {}\reflex cluster catalogue is well suited for studying the
large--scale structure of the Universe. The detection of the clusters
is based on their X--ray flux, allowing the construction of a
flux--limited sample.  X--ray selected cluster catalogues are not
impaired by projection effects. Moreover, the flux--limit, together
with the well documented selection effects allows the extraction of
clean volume--limited samples.

We calculated Minkowski functionals of a series of volume--limited
samples, extracted from the {}\reflex cluster catalogue. 
The comparison with the MFs of Poisson distributed points revealed
similar features as detected in the Abell/ACO cluster sample
{}\citep{kerscher:abell}.  Although the number of clusters in the
samples is always less than one hundred, MFs allow for a sensitive and
discriminatory analysis. The stability of the results obtained from
this small number of points can be attributed to the additivity
property of the MFs, which served as a construction principle.

Our aim was the quantification of non--Gaussian features in the
large--scale distribution of clusters, therefore we first gave a
precise definition of a Gaussian point distribution, the
Gauss--Poisson process.  Contrary to a Gaussian random field,
constraints for number density and the two--point correlation function
arise.  In the smaller volume--limited samples L05 and L12 these
constraints are violated.  Hence, a Gauss--Poisson process with the
observed density and two--point correlation function does {\em not}
exist.  This is an indirect detection of higher--order correlation
functions. Higher--order correlation functions are needed to allow for
the increased variance. Clearly, the relevance of these higher--order
correlations has to be checked independently, e.g.\ using the MFs.
Due to the decreasing number density of galaxy clusters the deeper
volume--limited samples L20 and L30 comply with the constraints.  A
Gauss--Poisson process based on the observed correlation function
becomes feasible as a model. MFs summarize the influence of the
two--point correlations and higher--order correlations on the
morphology of large--scale structure. They include correlations of any
order in an integral way. We calculated the MFs for a general
correlated point set.  Detailed results were given for the
Gauss--Poisson process.
To quantify higher--order correlations in the cluster distribution we
compare the analytical known MFs known for the Gauss--Poisson process
with the actual observed MFs of the cluster distribution. Two of the
four MFs, the volume and the surface area, are consistent with the
Gaussian model.  However a clear detection of non--Gaussian features
at large scales was possible with the integral mean curvature and the
Euler characteristic.

The definition of the Gauss--Poisson process directly lead to a method
for simulating Gaussian point distributions. With such simulated point
distributions we performed a non--parametric Monte--Carlo test. The
main result is that we can exclude a Gauss--Poisson process as a
viable model for the distribution of galaxy clusters at the
significance level of 95\%. 

Non--Gaussian features seen in the distribution of galaxy clusters may
be already imprinted on the initial density field (see e.g.\
{}\citealt{linde:nongaussian}), or may be a result of topological
defects (see e.g.\ {}\citealt{shellard:clusters}).
We would like to point out that also explanations facilitating
Gaussian initial conditions are possible.  Introducing a threshold
and considering only peaks in a Gaussian density field
{}\citet{bardeen:gauss} could show that the point distribution of the
peaks has non zero higher--order correlations $\xi_n\ne0$ for $n>2$.
Still the importance of the higher--order correlations on large scales
comes as a surprise within this model.  On physical grounds, the peak
biasing picture may only serve as a first approximation.  Evolving a
Gaussian density field in time using the linear approximation leads to
larger and larger regions with a non--physical negative mass
density.  Only the non--linear evolution of the density field can
remedy these shortcomings, allowing on the one hand for high density
peaks with an over--density of several hundreds, and on the other hand
allowing voids with a negative density contrast always larger than
minus one.  At the peaks of this non--linear evolved density fields
one may assume the clusters to reside.
As already discussed in the introduction non--Gaussian features in the
large--scale distribution of mass, like walls and filaments, are
predicted both by the Zel'dovich and related approximations as well as
by $N$--body simulations, both based on Gaussian initial conditions.
These structures in the mass distribution, perhaps amplified by a
biasing mechanism, can be associated with the non--Gaussian structures
observed in the large--scale distribution of {}\reflex galaxy
clusters. Our results suggest that within these scenarios, using
Gaussian initial conditions, it is necessary to consider non--linear
models to describe the observed large--scale structures.

\subsection*{Acknowledgments}

We thank J{\"o}rg Retzlaff for help in generating
Fig.~\ref{fig:bobbel}.  MK acknowledges support from the NSF grant
AST~9802980 and from the {\em Sonderforschungsbereich 375 f{\"u}r
Astroteilchenphysik der DFG}.  KM acknowledges support from the DFG
grant ME1361/6-1.

%%%%%%%%%%%%%%%%%%%%%%%%
%\bibliographystyle{apj} 
%\bibliography{my}

\appendix
\section{The constraints}
\label{sect:derive-constraints}

In this appendix we  recall the derivation of the constraints on
the number density $\varrho$ and $\xi_2(r)$ for the Gauss--Poisson
process as presented by {}\citet{milne:further} (see also
{}\citealt{kerscher:constructing}).

Consider $k$ compact disjoint sets $A_j$, and let $n_j=N(A_j)$ be the
number of points inside $A_j$.  The probability generating function of
the $k$-dimensional random vector $(n_1,\cdots,n_k)$ is then
\begin{equation}
\label{eq:gen-function}
P_k(\bz)=P_k(z_1,\ldots,z_k)= \BE\bigg[\prod_{j=1}^k z_j^{n_j}\bigg] .
\end{equation}
Together with a continuity requirement the knowledge of {\em all}
finite--dimensional probability generating functions $P_k$ determines
the p.g.fl.\ $G[h]$ (see Eq.~\eqref{eq:pgfl-xin}) and the point
process completely (e.g.\ {}\citealt{westcott:probability}).  Setting
\begin{equation}
\label{eq:def-hstrich}
h(\bx) = 1-\sum_{j=1}^k (1-z_j)\Beins_{A_j}(\bx) ,
\end{equation}
one obtains the probability generating function of these
finite--dimensional distributions from the probability generating
functional of the point process: $P_k(\bz)=G[h]$.  Here
$\Beins_A(\bx)$ is the indicator--function of the set $A$, with
$\Beins_A(\bx)=1$ for $\bx\in{A}$ and zero otherwise.

Since $P_k(\bz)$ is a probability generating function of a random
vector, it is positive and monotonically increasing in each component
$z_i$, hence $\log P_k(\bz)$ is non--decreasing.  Inserting
Eq.~\eqref{eq:def-hstrich} into the probability generating functional
of the Gauss--Poisson process {}\eqref{eq:G-gauss} one immediately
obtains
\begin{multline}
\frac{\partial\log P_k(\bz)}{\partial z_l} =
\varrho|A_l| + \\
+\varrho^2\sum_{j=1}^k \int_{A_l}\rmd\bx\int_{A_j}\rmd\by\
\xi_2(|\bx-\by|)(z_j-1) \ge 0
\end{multline}
for any $z_j\ge0$, where $|A_l|$ is the volume of the set $A_l$.
We chose $z_j=1$ for all $j\ne i$ and set the remaining $z_i$ either
to $z_i=0$ or $z_i\gg1$. Then the following two constraints emerge:
\begin{align}
\label{eq:const1}
\frac{\varrho}{|A_i|}\ \int_{A_i}\rmd\bx\int_{A_j}\rmd\by\ \xi_2(|\bx-\by|) 
& \le1 ,\\
\label{eq:const2}
\int_{A_i}\rmd\bx\int_{A_j}\rmd\by\ \xi_2(|\bx-\by|) 
& \ge  0 ,
\end{align}
for any subset $A_i,A_j$ of $\BR^3$.  {}\citet{milne:further} showed
that these two conditions are necessary and sufficient for the
existence of the Gauss--Poisson process.

With $A_i$ as an infinitesimal volume element centered on the origin
and $A_j$ equal to some volume $A$ the first constraint
{}\eqref{eq:const1} implies the simplified constraint
{}\eqref{eq:const2-alternate}. Considering two volume elements
$A_i=\rmd V_i$ and $A_j=\rmd V_j$, then Eq.~\eqref{eq:const2} implies
Eq.~\eqref{eq:const1-alternate}.

\end{document}